\documentclass[twocolumn,showpacs,preprintnumbers,amsmath,amssymb,aps,prb,superscriptaddress]{revtex4-2}

\pdfoutput=1
\usepackage[pdftex]{graphicx}
\usepackage[english]{babel}
\usepackage{soul}
\usepackage{cleveref}
\usepackage{bbold}
\usepackage{mathtools}
\usepackage{multirow}
\usepackage{colortbl}
\definecolor{kugray5}{RGB}{224,224,224}
\usepackage[latin1]{inputenc}
\usepackage{diagbox}
\usepackage{mciteplus}
\usepackage{mathrsfs}

\usepackage{dcolumn}
\usepackage{bm}

\usepackage{color}
\usepackage{amstext}
\usepackage{braket}
\usepackage{mathdots}
\usepackage{tikz}
\usepackage{physics}
\usepackage{enumitem}
\usepackage{siunitx}

\usetikzlibrary{matrix,decorations.pathreplacing}

\begin{document}


\title{Exceptional horns in $n$-root graphene and Lieb photonic ring lattices}

\author{A. M. Marques}
\email{anselmomagalhaes@ua.pt}
\affiliation{Department of physics $\&$ i3N, University of Aveiro, 3810-193 Aveiro, Portugal}

\author{D. Viedma}
\affiliation{Departament de F\'isica, Universitat Aut\`onoma de Barcelona, 08193 Bellaterra, Spain}

\author{V. Ahufinger}
\affiliation{Departament de F\'isica, Universitat Aut\`onoma de Barcelona, 08193 Bellaterra, Spain}
\affiliation{ICFO -- Institut de Ciencies Fotoniques, The Barcelona Institute of Science and Technology, 08860 Castelldefels, Spain}

\author{R. G. Dias}
\affiliation{Department of physics $\&$ i3N, University of Aveiro, 3810-193 Aveiro, Portugal}


\begin{abstract}
We present a systematic construction of non-Hermitian tight-binding lattices whose Bloch spectra are 
$n$th roots of those of Hermitian parent two-dimensional (2D) lattices, namely graphene and the Lieb lattice. 
The $n$-roots of these models are constructed from connecting loop modules of unidirectional couplings whose geometrical arrangements match those of the corresponding parent system.
Their energy spectrum is shown to consist of $n$ rotated and equivalent branches in the complex energy plane, each matching the real spectrum of the parent model when raised to the $n$th power, together with extra zero-energy flat bands (FBs) accounted for by the generalized index theorem.
We show how the low-energy Dirac cones of the parent models translate, for an appropriate choice of phase configuration for the couplings of the $n$-root lattices, as what we call an ``exceptional horn'' appearing at each branch, with the central Dirac point (DP) converted into zero-energy exceptional points (EPs) of order $n$ or higher at high-symmetry momenta. 
These exceptional horns reflect the behavior of low-lying excitations that scale with momentum as $E\sim\vert \mathbf{q}\vert^{\frac{1}{n}}$, with $n\geq 3$, as opposed to the linear massless modes that characterize a Dirac cone.
Moreover, we derive analytic expressions for the associated Landau levels (LLs), whose energies scale with magnetic flux as $E\sim\phi^{\frac{1}{2n}}$.
For the case of the $n$-root Lieb lattice, the zeroth LL is shown to be exceptional.
These results are analytically derived for both $n$-root models and numerically demonstrated for certain values of $n$.
Finally, we propose a realistic photonic implementation based on coupled ring resonators with a split configuration of optical gain and loss. This distribution yields strongly asymmetric couplings nearing unidirectionality, and appropriate positioning of the rings allows for fine-tuning of the coupling phases. We also discuss the impact of experimental imperfections on the EP signatures, modeling them through tight-binding simulations. 
\end{abstract}


\maketitle

\section{Introduction}
Recently, a lot of attention has been paid to a new class of topological insulators (TIs) \cite{Araujo2021} whose spectral and topological properties are inherited from a parent model hidden in a given power of their Hamiltonians.
These can be further classified into three distinct categories.
One is that of square-root topological insulators (TIs) \cite{Arkinstall2017}, whose parent TI appears as a diagonal block in the squared Hamiltonian \cite{Kremer2020,Pelegri2019,Mizoguchi2020,Ezawa2020,Ke2020,Mizoguchi2021c,Yoshida2021,Lin2021,Wu2021,Bomantara2022,Zhang2022,Matsumoto2023,Geng2023,Roy2024,Zhao2025}, which have already been realized in a variety of platforms \cite{Yan2020,Song2020,Yan2021,Song2022,Cheng2022,Kang2023,Yan2023,Wu2023,Guo2023,Yan2024,Song2024}.
Another category, an extension of the first one, is that of $2^n$-root TIs \cite{Marques2021,Marques2021b,Dias2021,Marques2023,Geng2024,Dias2025}, whose parent TI is only revealed after a sequence of $n$ squaring operations to the Hamiltonian (with energy shifts applied after each operation). 
Models in this class have recently been implemented in photonic \cite{Wei2023} and acoustic \cite{Cui2023} lattices.
Finally, there is the non-Hermitian class of $n$-root TIs \cite{Zhou2022,Viedma2024}, with $n\geq 3$ an integer, whose parent TI, which can be Hermitian, is directly obtained from the $n$th power of the Hamiltonian.
The root operation enlarges the unit cell, introduces unidirectional hoppings, and creates multiple copies of the parent spectrum rotated in the complex plane \cite{Viedma2024}.

Non-Hermitian extensions of quantum and classical systems have revealed a wealth of phenomena absent in Hermitian physics, foremost among them EPs where eigenvalues and eigenvectors coalesce \cite{Miri2019}. 
EPs of order two have been observed in a variety of platforms, but recent theoretical advances demonstrate that different mechanisms can be used to generate EPs of any order \cite{Wang2019,Sayyad2022,Tschernig2022,Wiersig2025}. 
The strong spectral response of these EPs to perturbations have made them attractive for their possible use in sensing applications \cite{Hodaei2017,Wiersig2020,Mao2024}.

Here, we devise a route to high-order EPs that exploits the algebraic structure of $n$-root tight-binding models. 
Specifically, we take the parent models to be graphene and the Lieb lattice, both of which characterized by low-energy Dirac-type physics.
The resulting band structures of their $n$-root versions, constructed with loop modules of unidirectional couplings, display a mixture of DPs and EPs whose order is dictated by the eigenvalue of the parent model at the expansion point. 
When the parent exhibits a finite energy DP at a high-symmetry momentum, the $n$-root lattice hosts an $n$-fold set of DPs at the same momentum in the complex energy plane. 
Conversely, if the DP of parent band is pushed to zero energy, which we show that can be realized by imposing appropriate phase configurations on the couplings, the $n$-root lattice displays EPs of order $n$ (or higher when FB eigenvectors also coalesce).
The dispersion relation in the vicinity of these EPs is found to scale as $E\sim\vert \mathbf{q}\vert^{\frac{1}{n}}$.
When a uniform transverse magnetic is applied to the $n$-root lattice, the magnetic flux dependency of the energy of the LLs is also found to go with the $n$th root of that of the parent model.

The presence of FBs in $n$-partite lattices with sublattice imbalance \cite{Marques2022}, such as the $n$-root models studied here, further enriches the EP structure. 
The generalized index theorem for non-Hermitian $n$-partite lattices guarantees a number of zero-energy states bounded from below by the sublattice imbalance. 
In an $n$-root lattice these FB eigenvectors merge with the EP eigenvectors, raising the EP order beyond $n$. 
However, they remain flat  under the momentum perturbation $\mathbf{q}$, not affecting the scaling of the dispersive bands.

For the experimental realization of the $n$-root lattices, the key factor is the implementation of unidirectional or non-Hermitian couplings. This challenge has been tackled in recent years, leading to experimental demonstrations in multiple platforms, such as acoustic, photonic and dipole resonators \cite{Wang2019,Zhang2021,Gu2022,Gao2022,Liu2022,Gao2023}, electrical circuits \cite{Hofmann2019,Helbig2020,Liu2021,Zou2021}, optical fibers \cite{Weidemann2020,Weidemann2022} and modulated waveguides \cite{Qin2020,Zheng2022,Ke2023}. Focusing on photonic ring resonators, the required bias is achieved by employing a modulation of optical gain and loss in engineered link resonators \cite{Longhi2015,Lin2021,Viedma2024}, which effectively enhance the coupling in one direction while suppressing it in the opposite one, approximately reaching unidirectionality. Here, we extend this platform to implement cubic-root versions of graphene and of the Lieb lattice by choosing the appropriate configurations for the link rings in each case. Resonator systems are also very well suited for the precise tuning of coupling strengths and phases \cite{Hafezi2011,Hafezi2013,Mittal2014}, thus allowing to implement the phase configuration that introduces a controlled energy downshift. With this, one is able to generate EPs at high-symmetry points of the Brillouin-zone.  
By combining analytic perturbation theory with finite-element simulations, we elucidate the origin of the complex dispersion $E\sim\vert \mathbf{q}\vert^{\frac{1}{n}}$ at low-energies, the formation of exceptional Landau levels under a synthetic magnetic field, and the Hofstadter butterfly patterns that arise in the presence of a uniform flux.

The rest of the paper is organized as follows.
In Sec.~\ref{sec:nrootgraph}, we introduce the $n$-root graphene model and determine the phase configuration for which the Dirac cones of the complex energy spectrum are pushed to zero energy and become exceptional horns.
The behavior of the model under a uniform perpendicular magnetic field is analyzed and the LLs are computed around the EP.
Sec.~\ref{sec:nrootlieb} follows the same structure as the previous section, only applied now to the $n$-root Lieb lattice.
Exact analytical solutions for the LLs associated with the branch FBs are also provided, and their exceptionality discussed.
Sec.~\ref{sec:photonic} describes how photonic ring systems with modulated gains and losses can implement close to ideal $n$-root lattices for low $n$ values, while offering a high degree of parametric control. 
Both for concreteness and validation purposes, we focus on the photonic 3-root graphene model and compare the numerical eigenspectrum with that of the theoretical model.
Possible causes of deviations in the numerical results are identified and discussed.
Finally, we present our conclusions in Sec.~\ref{sec:conclusions}.

	

\section{$n$-root graphene}
\label{sec:nrootgraph}

\begin{figure}[ht]
	
	\begin{centering}
		\includegraphics[width=0.48 \textwidth]{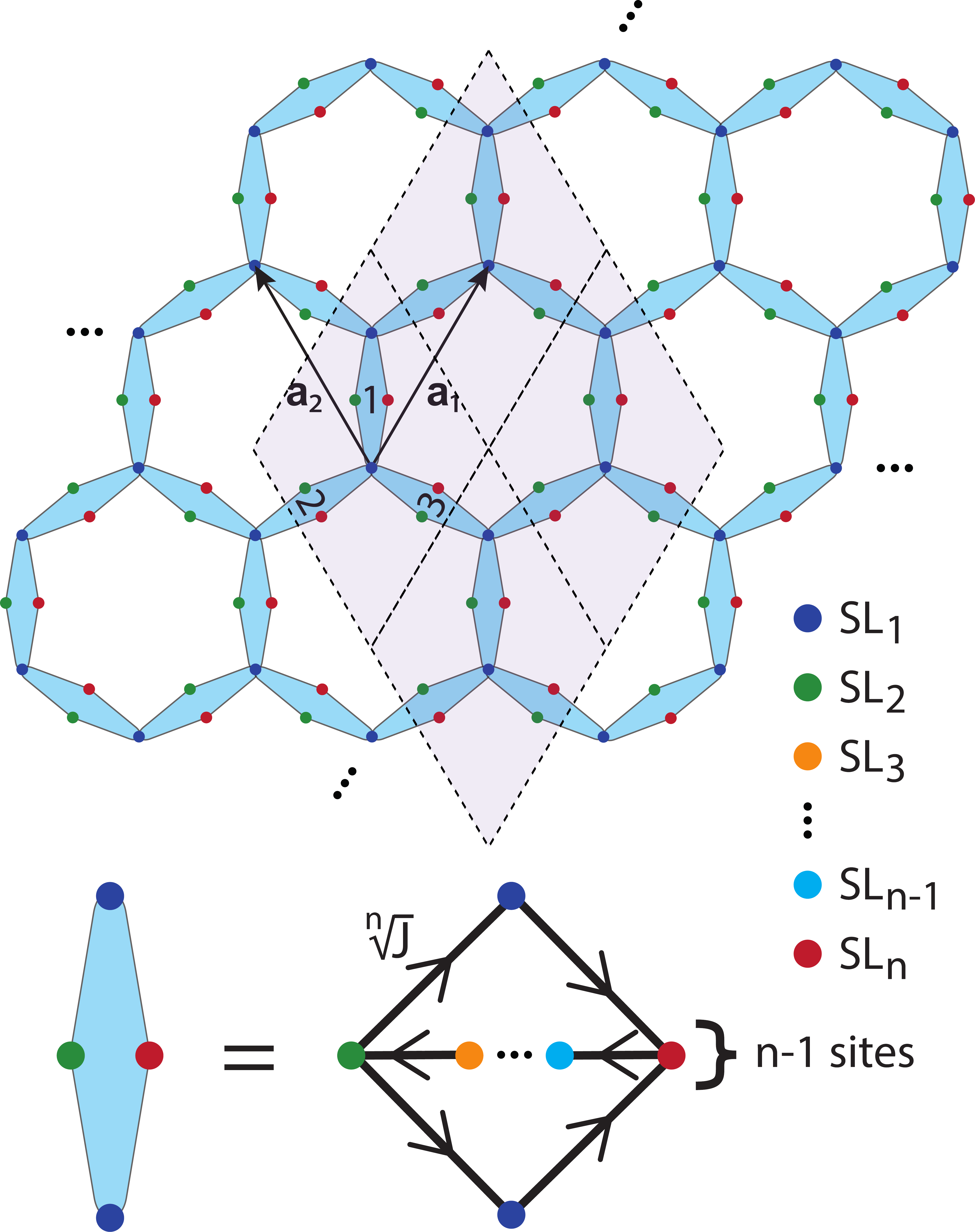}
		\par\end{centering}
	\caption{Illustration of the $n$-root graphene model. 
		Each dashed delimited shaded diamond encloses a unit cell. The primitive vectors are $\mathbf{a_1}=a\left(\frac{1}{2},\frac{\sqrt{3}}{2}\right)$ and $\mathbf{a_2}=a\left(-\frac{1}{2},\frac{\sqrt{3}}{2}\right)$, with $a\equiv 1$ the lattice constant.
	The general form of the loop modules is depicted at the bottom, with the arrows indicating the direction of the unidirectional couplings of magnitude $\sqrt[n]{J}$. The color scheme at the right indicates the SL to which each site belongs.}
	\label{fig:nrootgraphene}
\end{figure}

As depicted in Fig.~\ref{fig:nrootgraphene}, the $n$-root graphene model, with $n\geq3$, can be constructed by substituting the nearest-neighbor (NN) hoppings of the hexagonal lattice with loop modules made of unidirectional couplings \cite{Viedma2024}. Each of these modules comprises two sites of the blue sublattice (SL), corresponding to the SL of the parent graphene model, that connect to each other via $n$ unidirectional hopping terms passing through $n-1$ intermediate sites, forming a closed loop structure.
We assume at this point that a uniform hopping amplitude in the loop of  $\sqrt[n]{J}$, with $J\geq 0$, such that no phases are present.
The $n$-root model is $n$-partite \cite{Marques2022},  meaning it is constituted by $n$ SLs.
Sites belonging to different SLs are depicted with different colors in Fig.~\ref{fig:nrootgraphene}, with the $j$th SL labeled as SL$_j$, with $j=1,2,\dots,n$.
Within a unit cell, each SL has three sites, one per loop module, except for the blue SL$_1$, which only has two sites.

\begin{figure}[ht]
	
	\begin{centering}
		\includegraphics[width=0.48 \textwidth]{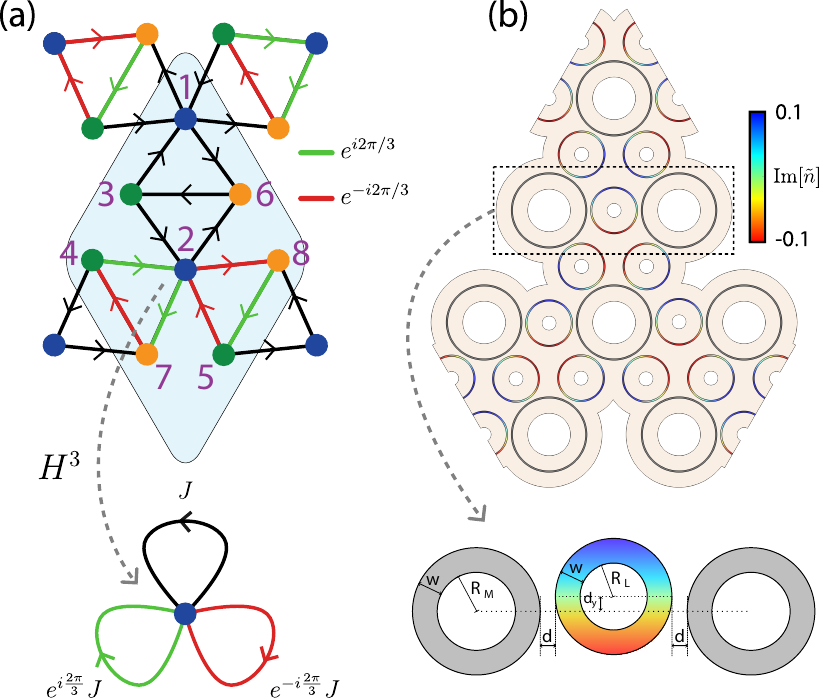}
		\par\end{centering}
	\caption{(a) Unit cell of the 3-root graphene model with a modified hopping phase configuration, with green and red unidirectional hoppings of magnitude $\sqrt[3]{J}$ carrying the respective phase factor indicated at the right. 
	The onsite energy terms appearing in the cubed Hamiltonian for the selected blue site are depicted as self-loops at the bottom right.
	(b) Photonic ring implementation of the unit cell shaded in blue in (a). The gray rings are neutral, and constitute the main rings of the effective lattice. For the antiresonant link rings, the sine-like distribution of the imaginary part of the refractive index is represented following the color scale to the right. The links corresponding to the phase terms are displaced perpendicularly from the line joining their main rings by a distance $d_y$.}
	\label{fig:3rootgrapheucell}
\end{figure}

The unit cell basis is written as
\begin{equation}
	\mathcal{B}:=\left\{\{\text{SL}_1\},\{\text{SL}_2\},\dots,\{\text{SL}_n\}\right\},
	\label{eq:slbasis}
\end{equation}
where $\{\text{SL}_1\}$ corresponds to the blue sites of SL$_1$ ordered by top first and bottom second, while $\{\text{SL}_i\}$, with $i=2,3,\dots,n$, corresponds to the sites of SL$_i$ ordered by loop module number (see the highlighted unit cell in Fig.~\ref{fig:nrootgraphene}).
In this basis, the bulk Hamiltonian takes the form
\begin{eqnarray}
	H(\mathbf{\mathbf{k}})&=&
	\begin{pmatrix}
		&h_1&&&
		\\
		&&h_2&&
		\\
		&&&\ddots&
		\\
		&&&&h_{n-1}
		\\
		h_n&&&&
	\end{pmatrix},
	\label{eq:hamiltnroot}
	\\
	h_1&=&h_n^\dagger=\sqrt[n]{J}
	\begin{pmatrix}
		1&e^{i\mathbf{k}\cdot\mathbf{a_1}}&e^{i\mathbf{k}\cdot\mathbf{a_2}}
		\\
		1&1&1
	\end{pmatrix},
	\label{eq:h1andn}
	\\
	h_l&=&\sqrt[n]{J}\mathbb{1}_3, \ \ l=2,3,\dots,n-1,
	\end{eqnarray}  
where the entries not shown are zeros and $\mathbb{1}_m$ is the identity matrix of size $m$.
For concreteness, the Hamiltonian for the $n=3$ case, with the site ordering within the unit cell of Fig.~\ref{fig:3rootgrapheucell}(a), in agreement with \eqref{eq:slbasis}, is written as
\begin{equation}
	H(\mathbf{k})=\sqrt[3]{J}
	\begin{pmatrix}
		0&0&1&e^{i\mathbf{k}\cdot\mathbf{a_1}}&e^{i\mathbf{k}\cdot\mathbf{a_2}}&0&0&0
		\\
		0&0&1&1&1&0&0&0
		\\
		0&0&0&0&0&1&0&0
		\\
		0&0&0&0&0&0&1&0
		\\
        0&0&0&0&0&0&0&1		
        \\
        1&1&0&0&0&0&0&0
        \\
        e^{-i\mathbf{k}\cdot\mathbf{a_1}}&1&0&0&0&0&0&0
        \\
        e^{-i\mathbf{k}\cdot\mathbf{a_2}}&1&0&0&0&0&0&0
	\end{pmatrix}.
	\end{equation}

The presence of the generalized chiral symmetry \cite{Marques2022}, defined as $C_n H(\mathbf{k})C_n^{-1}=\omega_n^{-1} H(\mathbf{k})$, with $\omega_n=e^{i\frac{2\pi}{n}}$ and
\begin{equation}
	C_n=\text{diag}(\mathbb{1}_{2},\omega_n\mathbb{1}_{3},\dots,\omega_n^{n-2}\mathbb{1}_{3},\omega_n^{n-1}\mathbb{1}_{3}),
	\label{eq:genchiral}
\end{equation} 
imposes that every \textit{finite} energy eigenvalue $E_\alpha(\mathbf{k})$ of $H(\mathbf{k})$ comes in $n$-tuples of the form $\left\{\omega_n^p E_\alpha(\mathbf{k})\right\}$, with $p=0,1,\dots,n-1$.
The $n$th power of the Hamiltonian has a block diagonal form that decouples the SLs,
\begin{eqnarray}
	H^n(\mathbf{k})&=&\text{diag}\left[H_1(\mathbf{k}),H_2(\mathbf{k}),\dots,H_n(\mathbf{k})\right],
	\label{eq:nroothamiltgraph}
	\\
	H_j\mathbf{k})&=&h_jh_{j+1}\dots h_{j+n-1},
	\label{eq:diagblocks}
\end{eqnarray}
with $n+1\to 1$ from  the boundary condition.
In particular, the first block describes the bulk Hamiltonian of graphene with a global energy shift,
\begin{eqnarray}
	H_1(\mathbf{k})&=&3J\mathbb{1}_2+H_{\text{graphene}}(\mathbf{k}),
	\label{eq:h1block}
	\\
	H_{\text{graphene}}(\mathbf{k})&=&J
	\begin{pmatrix}
	0&f(\mathbf{k})
	\\
	f^*(\mathbf{k})&0
	\end{pmatrix},
\end{eqnarray}
with $f(\mathbf{k})=1+e^{i\mathbf{k}\cdot\mathbf{a_1}}+e^{i\mathbf{k}\cdot\mathbf{a_2}}$ .
Hermiticity is recovered for $H_1(\mathbf{k})$, as its hopping terms of magnitude $J$ are now fully reciprocal.
Inspection of the loop module depicted at the bottom of Fig.~\ref{fig:nrootgraphene} reveals the origin of this reciprocity.
For the two blue sites of SL$_1$, there is a sequence of $n$ unidirectional hopping processes of magnitude $\sqrt[n]{J}$ that connects one to the other. 
Their product, obtained when $H(\mathbf{k})$ is raised to the $n$th power, generates the reciprocal $J$ hopping term between the blue sites of SL$_1$. 
A similar reasoning explains the global energy shift in (\ref{eq:h1block}) of $3J$.
Each loop module can be decomposed into two triangular subloops for which a particle at an SL$_1$ site returns to the same site after $n$ unidirectional hoppings.
Since each SL$_1$ site connects in this way to itself through three such subloops (see Fig.~\ref{fig:nrootgraphene}), a self-energy term of $3J$ appears in $H_1(\mathbf{k})$.

Diagonalization of $H_1(\mathbf{k})$ in (\ref{eq:h1block}), setting $a\equiv 1$ here and everywhere below, yields the following energy bands,
\begin{multline}
	E_{1,\pm}(\mathbf{k})=3J\pm J\left|f(\mathbf{k})\right|  
	\\
	=3J\pm J\sqrt{3+2\cos(k_x)+4\cos(\frac{k_x}{2})\cos(\frac{\sqrt{3}}{2}k_y)},
	\label{eq:ebandsgraph}
\end{multline}
which are shared by all other diagonal blocks in (\ref{eq:nroothamiltgraph}) \cite{Marques2022}.
\begin{figure*}[ht]
	\begin{centering}
		\includegraphics[width=0.97 \textwidth]{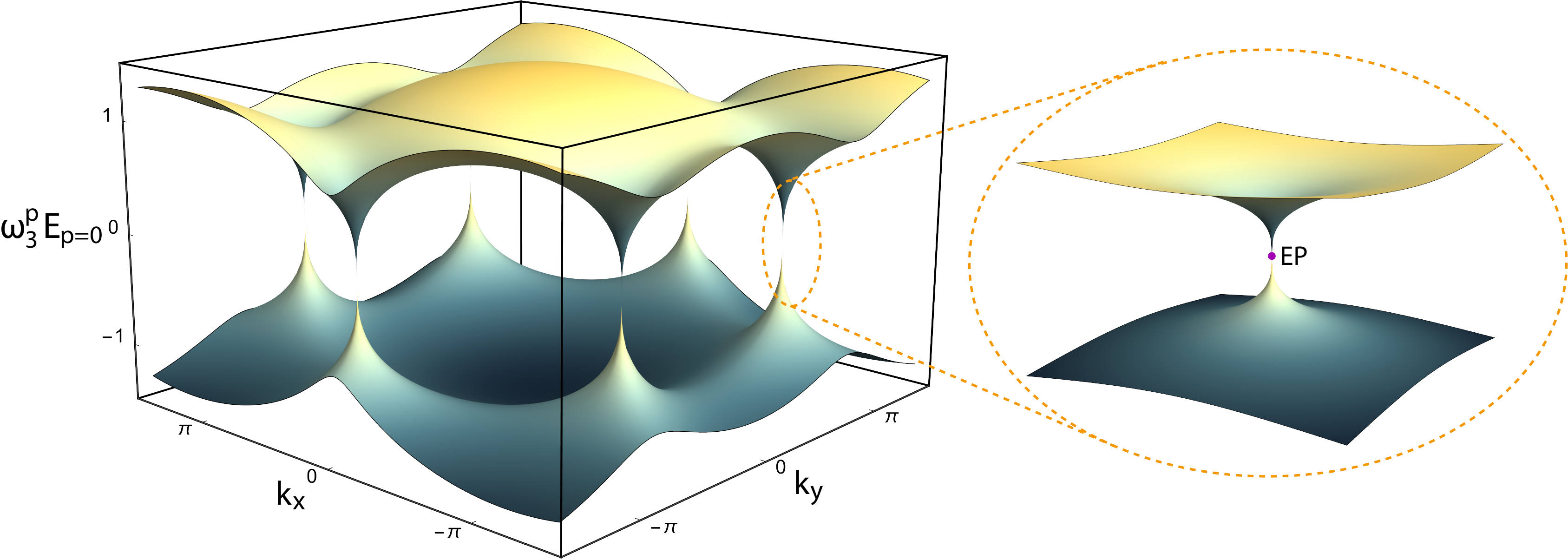}
		\par\end{centering}
	\caption{Bulk complex energy spectrum of branch $p=0,1,2$, with $E_{p=0}$ standing for the energies of the real zero branch, for the $3$-root graphene model, with $\omega_3^p=e^{i\frac{2\pi}{3}p}$ and $J=1$. A zoomed plot of the exceptional horn delimited by the dashed orange oval is shown at the right, where the zero-energy EP is highlighted.}
	\label{fig:3rootgraphene}
\end{figure*}
The complex energy spectrum of the $n$-root graphene model can be directly extracted from (\ref{eq:ebandsgraph}) as
\begin{equation}
E_{\pm}^{(p)}(\mathbf{k})=\omega_n^{p}E_{1,\pm}^{\frac{1}{n}}(\mathbf{k}),	
\label{eq:ebandsnrootgraph}
\end{equation}
where $p=0,1,\dots,n-1$ defines the \textit{energy branch}.
The spectral invariance under $\omega_n$ rotations is a consequence of the built-in $C_n$-symmetry of the $n$-root architecture.
It is this rotational invariance that imposes a branch structure to the finite energy spectrum in the complex plane.
The corresponding right eigenvectors for $E_{\pm}^{(p)}(\mathbf{k})\neq 0$ can be written as \cite{Andrade2024,Viedma2024,Gohsrich2024}
\begin{equation}
	\ket{\psi_{\pm}^{(p)}(\mathbf{k})}=\frac{1}{\sqrt{n}}
	\begin{pmatrix}
		\ket{\psi_{1,\pm}(\mathbf{k})}
		\\
		\omega_n^p E_\pm^{-\frac{n-1}{n}}h_2h_3\dots h_n\ket{\psi_{1,\pm}(\mathbf{k})}
		\\
		\omega_n^{2p} E_\pm^{-\frac{n-2}{n}}h_3\dots h_n\ket{\psi_{1,\pm}(\mathbf{k})}
		\\
		\vdots
		\\
		\omega_n^{(n-2)p} E_\pm^{-\frac{2}{n}}h_{n-1} h_n\ket{\psi_{1,\pm}(\mathbf{k})}
		\\
		\omega_n^{(n-1)p} E_\pm^{-\frac{1}{n}} h_n\ket{\psi_{1,\pm}(\mathbf{k})}
	\end{pmatrix},
	\label{eq:vecsnrootgraph}
\end{equation}
where $\ket{\psi_{1,\pm}(\mathbf{k})}$ is the eigenstate corresponding to $E_{1,\pm}(\mathbf{k})$ in (\ref{eq:ebandsgraph}).
Biorthogonality imposes the normalization condition $\braket{\tilde{\psi}_{\sigma\prime}^{(p\prime)}(\mathbf{k})}{\psi_\sigma^{(p)}(\mathbf{k})}=\delta_{\sigma,\sigma\prime}\delta_{p,p\prime}$, where $\sigma,\sigma\prime=\pm$ and $\ket{\tilde{\psi}_{\sigma\prime}^{(p\prime)}(\mathbf{k})}$ is the left eigenvector for band $\sigma\prime$ of branch $p\prime$ at the $\mathbf{k}$ momentum, satisfying $H^\dagger(\mathbf{k})\ket{\tilde{\psi}_{\sigma\prime}^{(p\prime)}(\mathbf{k})}=E_{\sigma\prime}^{*(p\prime)}(\mathbf{k})\ket{\tilde{\psi}_{\sigma\prime}^{(p\prime)}(\mathbf{k})}$.
From the generalized index theorem \cite{Marques2022}, there are $n-1$ extra flat bands coming from sublattice imbalance of each SL$_{j>1}$ with SL$_1$ that complete the spectrum. 
There is, in particular, an $n$-tuple of Dirac points (DPs) at the high-symmetry points $K=\left(\frac{4\pi}{3},0\right)$ and $K^\prime=\left(\frac{2\pi}{3},\frac{2\pi}{\sqrt{3}}\right)$, where $f\left(K^{(\prime)}\right)=0$,
\begin{equation}
	E_{\text{DP}}^{(p)}=E_{\pm}^{(p)}(K^{(\prime)})=\omega_n^{p}\left(3J\right)^{\frac{1}{n}}.
\end{equation}
These are non-degenerate complex energy DPs that exist at the $K$ and $K\prime$ points of each $p$ branch.
The expansion of $H(\mathbf{k})$ in (\ref{eq:hamiltnroot}) to first-order around $K$, with $\mathbf{q}:=\mathbf{k}-K$, only affects $h_1$ and $h_n$ in (\ref{eq:h1andn}), leading to
\begin{equation}
	h_n=h_1^\dagger\approx \sqrt[n]{J}
	\begin{pmatrix}
		1&1
		\\
		1&e^{-i\frac{2\pi}{3}}\left(1-\frac{i}{2}q_x-\frac{\sqrt{3}}{2}iq_y\right)
		\\
		1&e^{i\frac{2\pi}{3}}\left(1+\frac{i}{2}q_x-\frac{\sqrt{3}}{2}iq_y\right)
	\end{pmatrix}.
	\label{eq:hqgraphcomponents}
\end{equation}
The first diagonal block of $H^n(\mathbf{q})$ yields, to leading order in $\mathbf{q}$, the low-energy Hamiltonian of graphene with a global energy shift,
\begin{equation}
	H_1(\mathbf{q})\approx 3J\mathbb{1}_2+v_{_F}\left(q_x\sigma_x+q_y\sigma_y\right),
	\label{eq:lowenergygraph}
\end{equation}
where $v_{_F}=\frac{\sqrt{3}}{2}J$ is the Fermi velocity and $\sigma_x$ ($\sigma_y$) is the $x$ ($y$) Pauli matrix.
\begin{figure}[th]
	
	\begin{centering}
		\includegraphics[width=0.48 \textwidth]{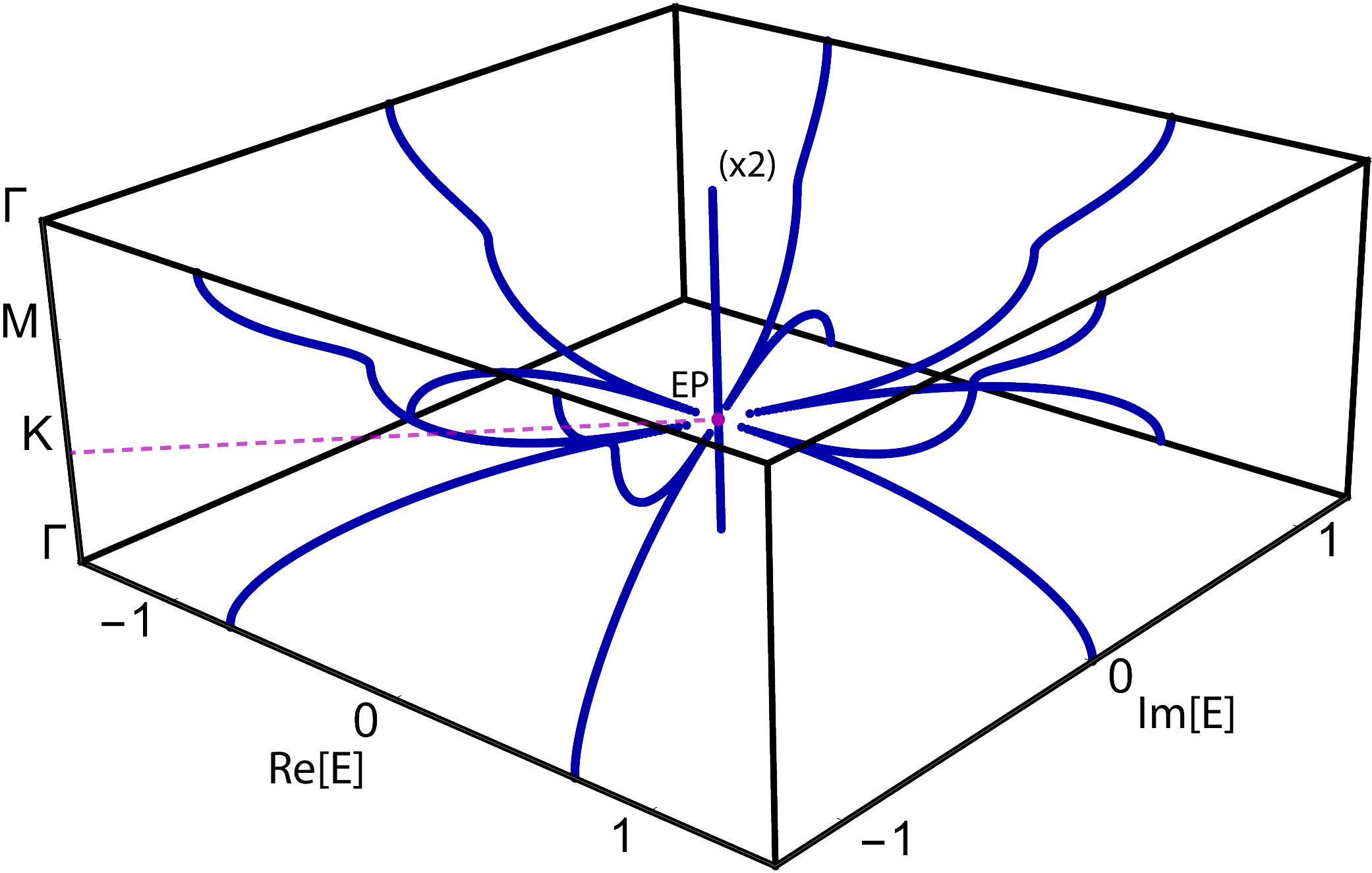}
		\par\end{centering}
	\caption{Bulk complex energy spectrum of the 3-root graphene along the high-symmetry line of the Brillouin zone (vertical axis), with $\Gamma=(0,0)$, $K=\left(\frac{4\pi}{3},0\right)$ and $M=\left(\pi,\frac{\pi}{\sqrt{3}}\right)$. The degeneracy of the zero-energy FB is indicated in parentheses. The purple dot at the center of the plot represents an EP at momentum $K$ and $E=0$. The small branch energy gaps around the EP are a numerical artifact.}
	\label{fig:enerhighsymgraph}
\end{figure}
Its energy spectrum is given by $E_\pm(\mathbf{q})=3J\pm v_{_F}|\mathbf{q}|$, which also holds for the other valley centered at $K^\prime$. As detailed in case $(i)$ of App.~\ref{app:perturbtheory}, the complex energy spectrum of  $H(\mathbf{q})$ is obtained from the linear expansion of $E_\pm^{\frac{1}{n}}(\mathbf{q})$, yielding
\begin{equation}
	E_{\pm}^{(p)}(\mathbf{q})\approx w_n^p\left[(3J)^{\frac{1}{n}}\pm v_{_F}^\prime |\mathbf{q}|\right],
\end{equation}
with a renormalized Fermi velocity
\begin{equation}
v_{_F}^\prime=\frac{1}{n}(3J)^{\frac{1-n}{n}}v_{_F}.
\end{equation}
This solution holds when a finite global shift is present in $H_1(\mathbf{q})$ [see the identity proportional term in (\ref{eq:lowenergygraph})].
In the next subsection, we will analyze the case where this global shift is absent.

\subsection{Turning Dirac points into exceptional points}

Suppose there is a way to get rid of the $E_0=3J$ energy shift in $E_\pm(\mathbf{q})$.
Then case (ii) of App.~\ref{app:perturbtheory} applies, and the spectrum of $H(\mathbf{q})$ directly becomes $E_{\pm}^{(p)}(\mathbf{q})=w_n^p E_\pm^{\frac{1}{n}}(\mathbf{q})$ which, after relabeling the bands to define each branch as group of two bands with symmetric energies, reads as
\begin{equation}
	E_{\pm}^{(p)}(\mathbf{q})=\pm w_n^{p/2} \left(v_{_F} |\mathbf{q}|\right)^{\frac{1}{n}},
	\label{eq:enernrootgraphep}
\end{equation}
corresponding to a \textit{sublinear} dispersion in $|\mathbf{q}|$, the hallmark of a system's response to a perturbation in the vicinity of an EP \cite{Miri2019}.
Mathematically, setting $E_0=0$ in $H_1(\mathbf{k})$ ensures that the root Hamiltonian $H(K^{(\prime)})$ becomes a nilpotent matrix (or a direct sum of nilpotent diagonal blocks), which can be converted into a Jordan chain through a similarity transformation. Since all eigenvalues collapse as $E_0\to0$ and the eigenvectors coalesce in a Jordan chain, this represents the formal definition of an EP at zero energy.

One way to downshift the global energy term to $E_0=0$ is by including sublattice dependent complex onsite potentials \cite{Martinez2024,Barnett2025}, corresponding to adding a complex term proportional to the generalized chiral symmetry $C_n$ to the Hamiltonian.
Here, we follow an alternative method that also retains $C_n$-symmetry and, therefore, the branch structure of the energy spectrum [see discussion below \eqref{eq:genchiral}]. This method is illustrated in Fig.~\ref{fig:3rootgrapheucell}(a) for $n=3$, where the unit cell depicted at the bottom of Fig.~\ref{fig:nrootgraphene} is altered by the introduction of a specific configuration of phases at the unidirectional hoppings $\sqrt[3]{J}$ and performing the corresponding changes to the $\{h_1,h_2,h_3\}$ elements of the Hamiltonian in (\ref{eq:hamiltnroot}).
After cubing the Hamiltonian of this 3-root graphene model, the onsite energy at the blue sites of SL$_1$, taking into account the phase accumulated along each of the three subloops that connect them to themselves [see bottom right of Fig.~\ref{fig:3rootgrapheucell}(a)], sums to $E_0=J\left(1+e^{i\frac{2\pi}{3}}+e^{-i\frac{2\pi}{3}}\right)=0$, as intended.
The branch energy spectrum, as defined in (\ref{eq:ebandsnrootgraph}), for this choice of phase configuration is shown in Fig.~\ref{fig:3rootgraphene}.
The absolute energy spectrum of each $p$ branch corresponds to the cubic-root of the absolute energy spectrum of graphene, but with the DP at the $K$ and $K^\prime$ points converted into an EP.
The zoomed low-energy region in Fig.~\ref{fig:3rootgraphene}  highlights the cubic-root dependence on $|\mathbf{q}|$ around the EPs of the 3-root graphene, in agreement with (\ref{eq:enernrootgraphep}).
The Dirac cones of the parent graphene model turn into what we label as \textit{exceptional horns} (in reference to the mathematical shape known as Gabriel's horn \cite{Coll2014}) in the $n$-root model, provided the phase configuration of the unidirectional couplings is such that $E_0=0$, leading to the low-energy expansion discussed in case (ii) of App.~\ref{app:perturbtheory}.
These exceptional horns are to be contrasted with the Dirac cones that appear in other non-Hermitian graphene based models \cite{Xue2020,Xie2025}, where the characteristic linear dispersion around the DP is preserved.
We clarify that the exceptional horn is not a new mathematical object, but rather a specific realization of a higher-order EP. 
Concretely, the term refers to the geometric shape of the complex dispersion in momentum space, defined by the sublinear scaling $E\sim\vert\mathbf{q}\vert^{1/n}$ away from the EP.
\begin{figure*}[ht]
	
	\begin{centering}
		\includegraphics[width=0.97 \textwidth]{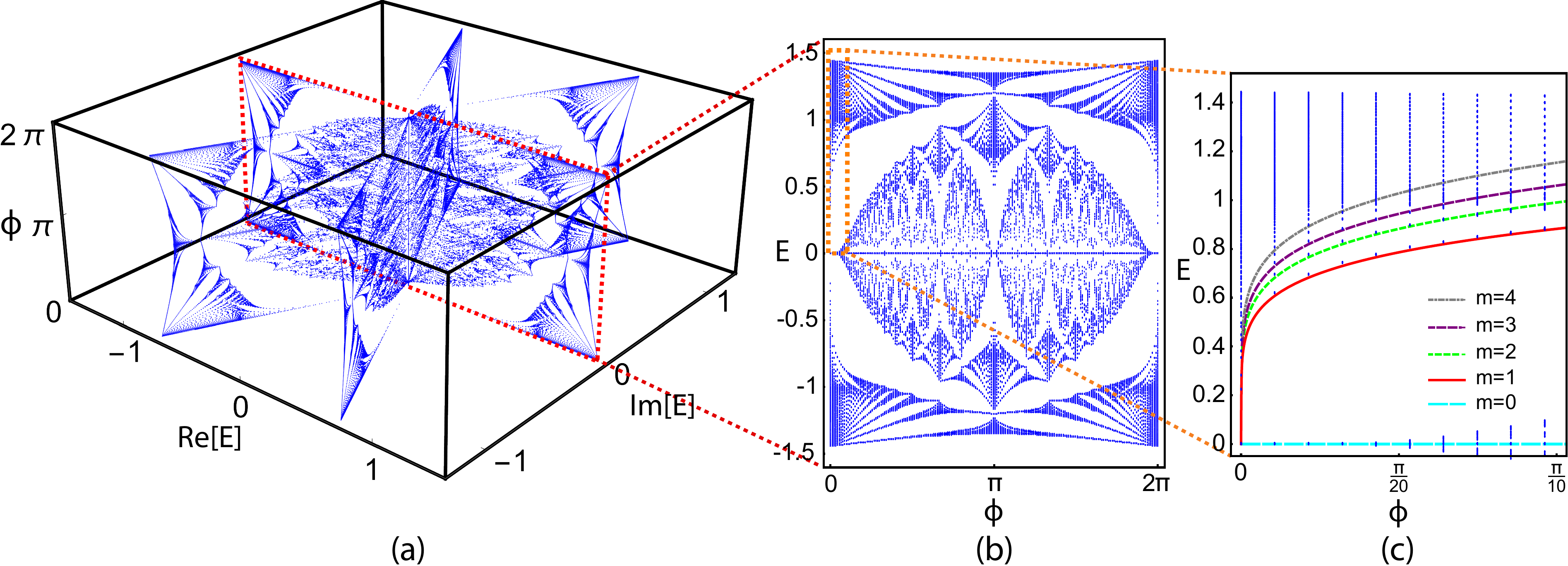}
		\par\end{centering}
	\caption{(a) Complex energy spectrum of the periodic 3-root graphene lattice as a function of $\phi=2\pi\frac{l}{q}$, with $l=0,1,\dots,q$ and $q=187$. (b) Zoomed plot of the $p=0$ real energy branch in (a) delimited by the dashed red box. (c) Zoomed plot of the low-flux region in (b) delimited by a dashed orange box. The curves of different colors and formats represent the energy $E_{+m}^{(0)}(\phi)$, given in (\ref{eq:enerllsnrootgraph}), of the corresponding LLs.}
	\label{fig:hofstspecgraph}
\end{figure*}
\subsection{Hofstadter spectrum and Landau levels of $n$-root graphene}
\label{subsec:hofstnrootgraph}

For the $n=3$ case considered here, and aside from the three energy branches, there are two extra zero-energy flat bands in the spectrum, originated from sublattice imbalance, as explained below (\ref{eq:ebandsnrootgraph}).
In Fig.~\ref{fig:enerhighsymgraph}, the full complex energy spectrum along the high-symmetry path of the Brillouin zone is depicted, where doubly degenerate zero-energy flat band states appear along the whole path.
At the exceptional $K$ point, all eight energy states coalesce at $E=0$, with six coming from the horns and two from the FBs. 
For a general $n$ value, the Jordan decomposition at the $K$ point reads as $H(K)=RJR^{-1}$,  with $J=J_n(0)\oplus J_{2n-1}(0)$, where $J_m(\lambda)$ is a Jordan block of size $m$ and eigenvalue $\lambda$, and $R$ is the transformation matrix whose columns are the generalized eigenvectors of the Jordan blocks.
The system is thus defective at the $K$ point with only two eigenvectors, corresponding to an EP$_n$ and to an EP$_{2n-1}$.
The eigenstates of the $n-1$ zero-energy flat bands coalesce into the eigenstate of the higher-order EP$_{2n-1}$ at the $K^\prime$ point.
However, since the perturbative term around $K^\prime$, namely $H(\mathbf{q})$ with components given in (\ref{eq:hqgraphcomponents}), retains the same sublattice structure of $H(K)$, the flat bands are still protected by the generalized index theorem \cite{Marques2022}, and they are not responsive to the perturbative parameter $\mathbf{q}$.
On the other hand, the dispersive bands of the energy branches have been shown in (\ref{eq:enernrootgraphep}) to scale with $\left|\mathbf{q}\right|^{\frac{1}{n}}$ around both EPs.
In particular, this means that, under the action of the $H(\mathbf{q})$ perturbation, the higher-order EP$_{2n-1}$ is decomposed into different cycles \cite{Moro1997,Ma1998,Demange2011}, one of which, of order $n$, encodes the response of the dispersive bands.
By writing the perturbation in the Jordan basis $\delta J(\mathbf{q})=R^{-1}H(\mathbf{q})R$,
the same EP phenomenology could be derived by  identifying the elements of $\delta J(\mathbf{q})$ that generate nontrivial perturbations  \cite{Jiang2020,Sayyad2022}, that is, those that cannot be expressed as elements of $\left[\delta S,J\right]$, with $\delta S$ an arbitrary perturbation matrix.

We turn now to the study of the $n$-root graphene lattice with the loop phase configuration studied in the previous subsection [see the $n=3$ case in Fig.~\ref{fig:3rootgrapheucell}(a)] in the presence of a uniform magnetic flux threading its plaquettes (white regions of the lattice depicted in Fig.~\ref{fig:nrootgraphene}).
We further impose that no flux goes through the loop modules, as their phase configuration is constrained by the requirement of canceling the global offset in the parent $H_1$ block.
These conditions on the phase configuration translate into a uniform flux per plaquette at the parent graphene model in SL$_1$, providing a direct correspondence with the well-known solutions for this case, of which we will make use to determine the Landau levels (LLs) of the $n$-root graphene model, as shown below.
We follow the standard procedure of square-shaping the $n$-root graphene lattice into a brick-wall lattice \cite{Jose2018} (whose hopping terms are replaced by loop modules) periodic now along the $x$ and $y$ directions, and insert a uniform reduced flux per plaquette $\phi=2\pi\frac{\Phi}{\Phi_0}$, where $\Phi$ is the magnetic flux and $\Phi_0$ the flux quantum.
Periodic boundary conditions (PBC) are applied  along both $x$ and $y$ directions, which can only be satisfied for rational flux values, $\phi=2\pi\frac{l}{q}$, where $l$ and $q$ are co-prime numbers.

In Fig.~\ref{fig:hofstspecgraph}(a), we plot the complex Hofstadter spectrum for $n=3$ under PBC, with $N_x=3$ ($N_y=q=187$) unit cells in the $x$-direction ($y$-direction), where the three-branch structure is preserved, since the flux does not break the generalized chiral symmetry $C_3$, meaning every finite energy state appears in triplets connected to each other through $\frac{2\pi}{3}$ rotations in the complex energy plane, as discussed below (\ref{eq:genchiral}) for the Bulk Hamiltonian. 
A zoom of the real-energy branch is shown in Fig.~\ref{fig:hofstspecgraph}(b), which cubes as a whole to the Hofstadter butterfly of graphene \cite{Rammal1985}.

In the previous section, the low-energy behavior of the $n$-root graphene model was extracted from the solutions of the expanded parent block $H_1(\mathbf{q})$ in (\ref{eq:lowenergygraph}), which led, after setting the diagonal term to zero, to the energy dispersion in (\ref{eq:enernrootgraphep}).
The same reasoning applies when determining the LLs of the $n$-root model around the EP \cite{Zhang2020}, \textit{i.e.}, we start directly from the solutions for the graphene parent block $H_1(\mathbf{q})$ (after removing the global energy shift of $3J$)  \cite{Goerbig2011} and apply $q_x\to i\sqrt{\frac{B}{2}}\left(b^\dagger - b\right)$ and  $q_y\to\sqrt{\frac{B}{2}}\left(b + b^\dagger\right)$ to the $h_n$ and $h_1$ elements in (\ref{eq:hqgraphcomponents}), where $b^{(\dagger)}$ is the lowering (raising) bosonic ladder operator, obeying $\left[b,b^\dagger\right]=1$ and $\left[b^{(\dagger)},b^{(\dagger)}\right]=0$, and acting on the number states as $b\ket{m}=\sqrt{m}\ket{m-1}$ and $b^\dagger\ket{m}=\sqrt{m+1}\ket{m+1}$.
The Landau gauge was assumed, with the vector potential given by $\mathbf{A}=xB\mathbf{\hat{y}}$, where $B$ is the magnitude of the uniform magnetic field.
The eigenvalues and eigenvectors obtained for the $K$-valley are written, respectively, as
\begin{eqnarray}
	E_{\pm m}(\phi)&=&\pm v_{_F}\sqrt{2m\phi},
	\label{eq:enerlandaugraph}
	\\
	\ket{\psi_{\pm m}}&=&\frac{1}{\sqrt{2}}
	\begin{pmatrix}
		\ket{m-1}
		\\
		\pm i\ket{m}
	\end{pmatrix},
	\label{eq:lleigsgraph}
\end{eqnarray}
where $m\in \mathbb{N}_0$, $\ket{m}$ is the number state of the 1D quantum harmonic oscillator, and natural units are assumed ($e,\hbar=1$, where $e$ is the electron charge and $\hbar$ is the reduced Planck constant). 
Setting $m=0$ and $\ket{-1}=0$, the zeroth LL has $E_{0}(\phi)=0$ and eigenvector $\psi_{0}=\left(0\ ,\ket{0}\right)^T$.

The energy of the LLs of the $n$-root graphene model can be directly determined from taking the $n$th root of (\ref{eq:enerlandaugraph}) and considering all branches in the complex energy spectrum,
\begin{equation}
	E_{\pm m}^{(p)}(\phi)=\pm\omega_n^{p/2}v_{_F}^\frac{1}{n}\left(2m\phi\right)^\frac{1}{2n}.
	\label{eq:enerllsnrootgraph}
\end{equation}
Fig.\ref{fig:hofstspecgraph}(c) zooms in on the low-flux region highlighted in Fig.\ref{fig:hofstspecgraph}(b), where the curves following the analytical expression in (\ref{eq:enerlandaugraph}) for the lowest LLs and $n=3$ are shown to be in good agreement with the numerical results.
In analogy with (\ref{eq:vecsnrootgraph}), an approximate solution for the right eigenvectors of the LLs, for finite flux, can be written as
\begin{equation}
	\ket{\psi_{\pm m}^{(p)}(\phi)}\approx\frac{1}{\sqrt{n}}
	\begin{pmatrix}
			\ket{\psi_{\pm m}}
			\\
			\omega_n^p E_{\pm m}^{-\frac{n-1}{n}}(\phi)h_2h_3\dots h_n\ket{\psi_{1,\pm m}}
			\\
			\omega_n^{2p} E_{\pm m}^{-\frac{n-2}{n}}(\phi)h_3\dots h_n\ket{\psi_{1,\pm m}}
			\\
			\vdots
			\\
			\omega_n^{(n-2)p} E_{\pm m}^{-\frac{2}{n}}(\phi)h_{n-1} h_n\ket{\psi_{1,\pm m}}
			\\
			\omega_n^{(n-1)p} E_{\pm m}^{-\frac{1}{n}}(\phi) h_n\ket{\psi_{1,\pm m}}
		\end{pmatrix}.
		\label{eq:vecsnrootgraphlandau}
\end{equation}
These eigenvectors are not exact since, as explained in App.~\ref{app:perturbtheory},  we only kept the terms to leading order in $|\mathbf{q}|$ in $H_1(\mathbf{q})$, from where $\ket{\psi_{1,\pm m}}$ is ultimately determined.
Compact analytical formulas for $\ket{\psi_{1,\pm m}}$, as in (\ref{eq:lleigsgraph}), cannot be found from the quantization of the exact form of $H_1(\mathbf{q})$, which includes terms $\propto |\mathbf{q}|^2$.
Note, however, that the approximate eigenstates in (\ref{eq:vecsnrootgraphlandau}) are \textit{flux-dependent}, in contrast to the usual case.
\subsubsection{Open boundary conditions}

\begin{figure}[ht]
	
	\begin{centering}
		\includegraphics[width=0.48 \textwidth]{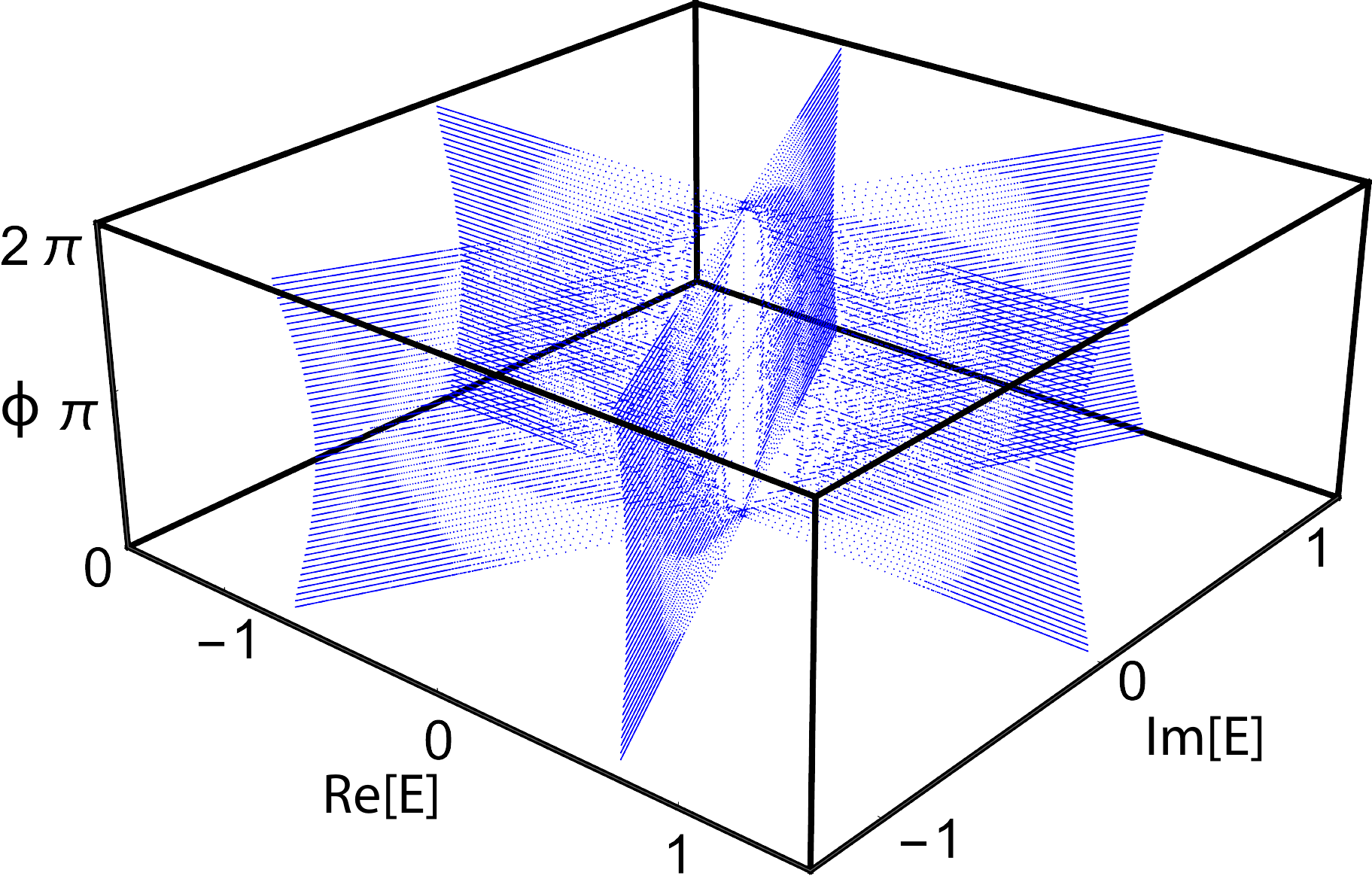}
		\par\end{centering}
	\caption{Complex energy spectrum of the fully open 3-root graphene lattice as a function of $\phi=2\pi\frac{l}{q}$, with $l=0,1,\dots,q$ and $q=51$, with $N_x=10$ and $N_y=q$, plux extra edge subloops to compensate for the lower connectivity of some boundary blue sites of SL$_1$.}
	\label{fig:hofstspecgraphobc}
\end{figure}

We briefly analyze here the Hofstadter spectrum of the $3$-root graphene under open boundary conditions (OBC).
The lattice considered has dimensions $N_x=10$ and $N_y=q=51$, plus added boundary sites \cite{Marques2021b} to ensure all blue sites of SL$_1$ are connected by three subloops in a way that keeps the phase configuration.
This is required to ensure that the cubed Hamiltonian corresponds to a clean open graphene lattice in the SL$_1$ block, that is, a zero onsite potential at all blue sites, including the boundary ones.
The resulting spectrum is shown in Fig.~\ref{fig:hofstspecgraphobc}.
It matches the one obtained for PBC in Fig.~\ref{fig:hofstspecgraph}(a), other than the boundary modes that traverse the gaps of the latter due to the open boundaries.
Notably, the skin effect is \textit{absent} for OBC (all bulk states remain extensive), even though the model is constructed with unidirectional hoppings.
Since the architecture of $n$-root constrains these unidirectional hoppings of constant magnitude to form loops that are both in-going to and out-going from their vertex blue sites [see Fig.~\ref{fig:3rootgrapheucell}(a)], there is no coalescence of bulk eigenstates to the boundaries, as no preferential direction is established across the lattice \cite{Marques2022}.
At the same time, the model remains exceptional at $E=0$ for $\phi=0\mod{2\pi}$, regardless of the boundary conditions since, for OBC, the parent Hamiltonian is still that of a graphene lattice (only now open) with no global energy shift, therefore its root versions retain the same exceptional spectral structure as for the case of periodic boundaries.

\section{$n$-root Lieb lattice}
\label{sec:nrootlieb}

The method developed in Sec.~\ref{sec:nrootgraph} for finding the $n$-root graphene model, based on substituting the hoppings of the parent hexagonal lattice with loop modules of unidirectional hoppings (see Fig.~\ref{fig:nrootgraphene}), can be generalized to other parent models.
In this section, we will take the parent model to be the Lieb lattice, which hosts a pseudospin-1 Dirac cone centered at the $\Gamma=(0,0)$ point in its bulk energy spectrum.
\begin{figure}[ht]
	
	\begin{centering}
		\includegraphics[width=0.46 \textwidth]{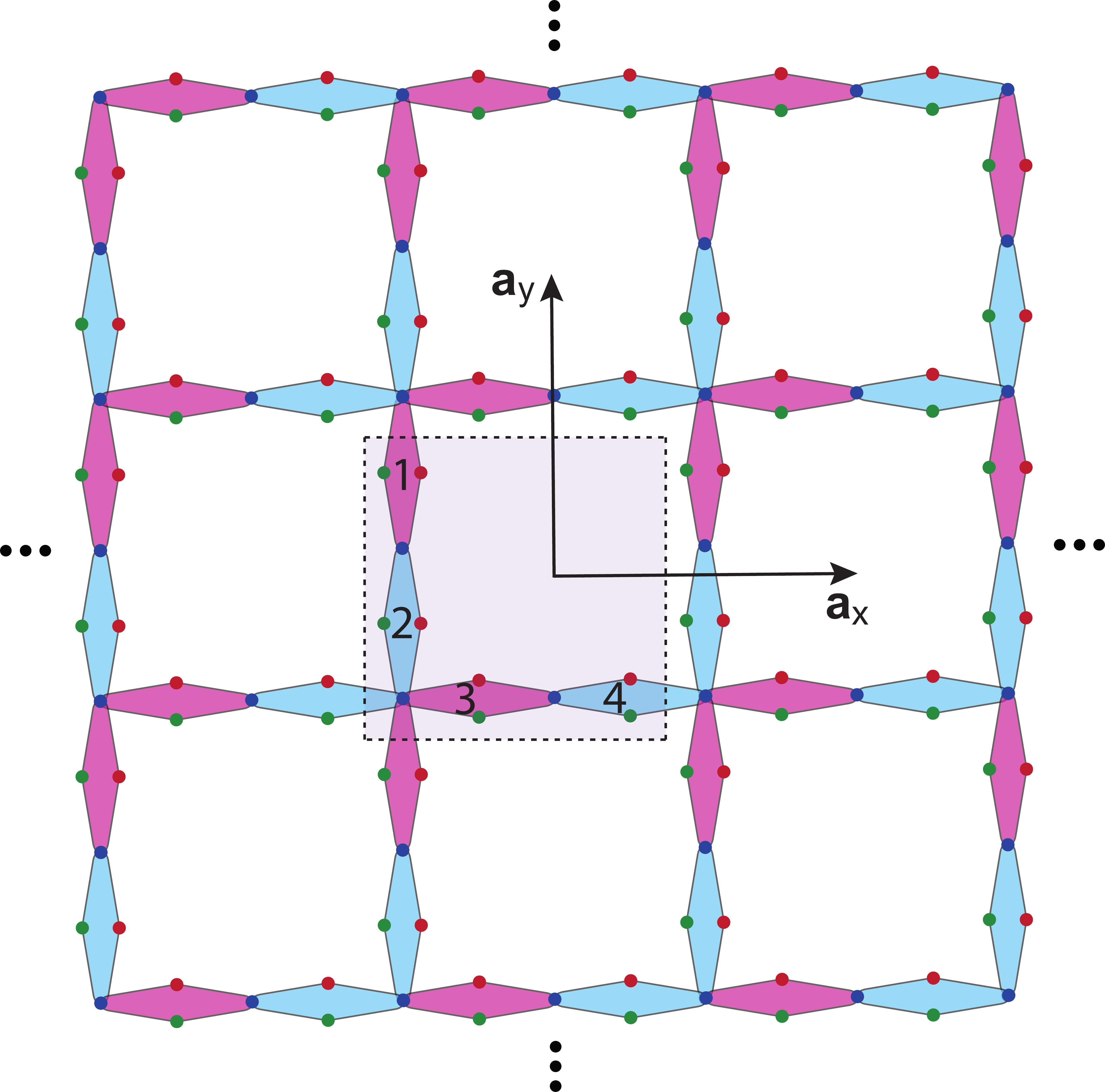}
		\par\end{centering}
	\caption{Illustration of the $n$-root Lieb lattice. Dashed limited square region encloses a unit cell. The primitive vectors are $\mathbf{a_x}=\left(a,0\right)$ and  $\mathbf{a_y}=\left(0,a\right)$, with $a\equiv 1$ the lattice constant. The blue and magenta loop modules have the configuration depicted at the bottom of Fig.~\ref{fig:nrootgraphene}, with the latter having an extra cumulative $\pi$ phase across the couplings between the red and green sites in the middle.}
	\label{fig:nrootlieb}
\end{figure}
Its $n$-root version is depicted in Fig.~\ref{fig:nrootlieb}.
Using the basis in (\ref{eq:slbasis}), where $\{\text{SL}_1\}$ corresponds to the blue sites of SL$_1$ ordered within each unit cell by corner first, middle top second and middle bottom third, while $\{\text{SL}_i\}$, with $i=2,3,\dots,n$, corresponds to the sites of SL$_i$ ordered by loop module number (see the highlighted unit cell in Fig.~\ref{fig:nrootlieb}), the bulk Hamiltonian of the $n$-root Lieb lattice, with the form of (\ref{eq:hamiltnroot}), has its elements now written as
\begin{eqnarray}
		h_1&=&h_n^\dagger=\sqrt[n]{J}
	\begin{pmatrix}
		e^{-ik_y}&1&1&e^{-ik_x}
		\\
		1&1&0&0
		\\
		0&0&1&1
	\end{pmatrix},
	\label{eq:h1andnlieb}
	\\
	h_l&=&\sqrt[n]{J}\text{diag}\left(e^{i\frac{\pi}{n-2}},1,e^{i\frac{\pi}{n-2}},1\right), l=2,3,\dots,n-1.
	\label{eq:hlnlieb}
\end{eqnarray}
The phase factors in (\ref{eq:hlnlieb}) are included at the unidirectional hoppings connecting the red and green sites of the magenta loop modules in Fig.~\ref{fig:nrootlieb}, that is, we consider a uniform distribution of phases at these hoppings that accumulates to $\pi$.
This choice of phase configuration was made so that the $H_1(\mathbf{k})$ diagonal block in (\ref{eq:nroothamiltgraph}) directly corresponds to the bulk Hamiltonian of the Lieb lattice $H_{\text{Lieb}}(\mathbf{k})$,
\begin{equation}
	H_1(\mathbf{k})=H_{\text{Lieb}}(\mathbf{k})=J
	\begin{pmatrix}
		0&1-e^{-ik_y}&e^{-ik_x}-1
		\\
		1-e^{ik_y}&0&0
		\\
		e^{ik_x}-1&0&0
	\end{pmatrix}.
	\label{eq:hamiltlieb}
\end{equation}
In other words, the phase configuration is such that there is no global energy shift in $H_1(\mathbf{k})$ [Hamiltonian for the blue sites in SL$_1$ appearing in $H^n(\mathbf{k})$].
This can be understood by noticing that each blue site in Fig.~\ref{fig:nrootlieb} is connected to itself through $n$ hopping processes by the same number of blue and magenta subloops.
Analogously to the case of 3-root graphene depicted at the bottom right of Fig.~\ref{fig:3rootgrapheucell}(a), the contributions of these subloops to the onsite energies in $H_1(\mathbf{k})$ cancel out: each blue subloop generates a $J$ onsite term, while each magenta subloop contributes with a $-J$ onsite term, where the negative sign is a consequence of the $\pi$ phase accumulated along the middle line of unidirectional hoppings.

Diagonalization of $H_1(\mathbf{k})$ in (\ref{eq:hamiltlieb}) yields two dispersive energy bands and a flat band,
\begin{eqnarray}
	E_{1,\pm}(\mathbf{k})&=&\pm J\sqrt{2\left(2+\cos k_x+\cos k_y\right)},
	\\
	E_{1,0}(\mathbf{k})&=&0,
\end{eqnarray}
from where the complex energy spectrum of the $n$-root Lieb lattice can be directly extracted as
\begin{equation}
	E_{\sigma}^{(p)}(\mathbf{k})=\omega_n^{p}E_{1,\sigma}^{\frac{1}{n}}(\mathbf{k}),	\ \ \ \sigma=0,\pm,
	\label{eq:ebandsnrootlieb}
\end{equation}
where $p=0,1,\dots,n-1$ is the branch index. 
Note that, in addition to the $n$ FBs for $\sigma=0$ in (\ref{eq:ebandsnrootlieb}), the full energy spectrum has $n-1$ extra zero-energy FBs coming from the imbalance between each SL and SL$_1$ \cite{Marques2022} (all SLs have four sites per unit cell, except for SL$_1$, which has three), yielding a total of $2n-1$ FBs.
\begin{figure}[ht]
	
	\begin{centering}
		\includegraphics[width=0.48 \textwidth]{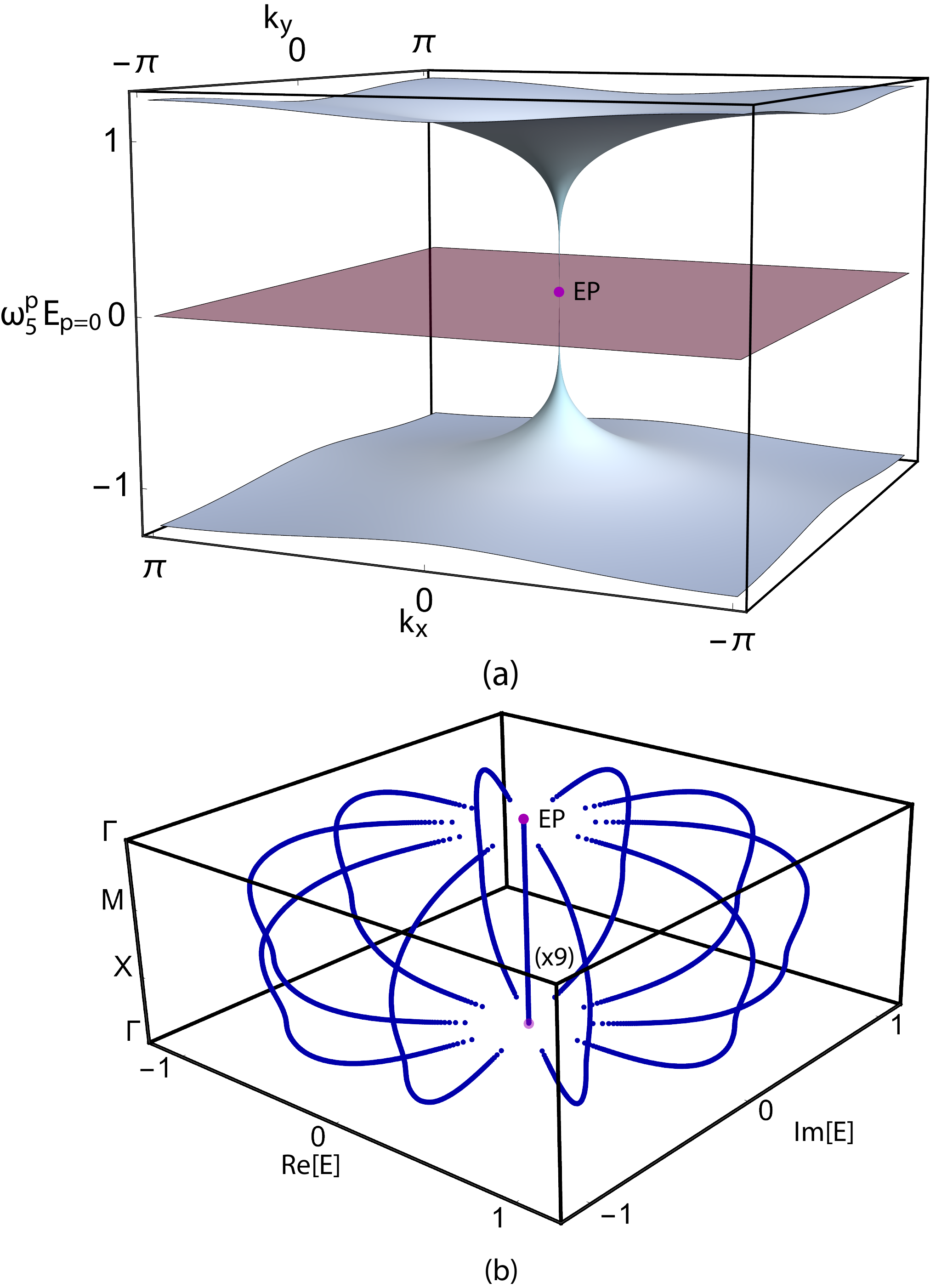}
		\par\end{centering}
	\caption{(a) Bulk complex energy spectrum of branch $p=0,1,2,3,4$, with $E_{p=0}$ standing for the energies of the real zero branch, for the $5$-root Lieb lattice, with $\omega_5^p=e^{i\frac{2\pi}{5}p}$ and $J=1$.
		(b) Bulk complex energy spectrum of the 5-root Lieb lattice along the high-symmetry line (vertical axis) of the Brillouin zone, with $\Gamma=(0,0)$, $X=\left(\pi,0\right)$ and $M=\left(\pi,\pi\right)$. The degeneracy of the zero-energy FB is indicated in parentheses. The purple dots represent the same EP at momentum $\Gamma$ and $E=0$. The branch energy gaps around the EP are a numerical artifact.}
	\label{fig:5rootliebspectrum}
\end{figure}

Fig.~\ref{fig:5rootliebspectrum}(a) shows the complex branch energy spectrum of the 5-root Lieb lattice, which is constituted of a symmetric exceptional horn touching the middle FB at the EP located at the $\Gamma$ point. 
Fig.~\ref{fig:5rootliebspectrum}(b) shows the full spectrum along the high-symmetry line in the Brillouin zone, where it can be seen that, at the exceptional $\Gamma$ point, all 19 energy states coalesce at $E=0$. 
For a general $n$ value, the Jordan decomposition at the $\Gamma$ point transforms the Hamiltonian into its Jordan block form, given by $J=J_1(0)\oplus J_{2n-1}(0)\oplus J_{2n-1}(0)$.
The system is thus defective at the $\Gamma$ point with only three eigenvectors, with one of them corresponding to a regular point and the other two to an EP$_{2n-1}$ each.
The low-energy expansion $H(\mathbf{q})$ around $\Gamma$ falls into case (ii) of App.~\ref{app:perturbtheory}, with the dispersive bands of each branch following (\ref{eq:enernrootgraphep}) with $v_{_F}=J$ for the Lieb lattice.
The FBs remain locked at zero-energy, as their number is fixed by the generalized index theorem \cite{Marques2022} relative both to $H(\mathbf{q})$ ($n-1$ FBs from SL imbalance) and $H_1(\mathbf{q})$ [low-energy expansion of the Lieb lattice has one FB that appears in the spectrum of the other diagonal blocks also, due to the isospectrality condition, which gives rise to $n$ FBs in the spectrum of $H(\mathbf{q})$].
It is also clear from $E_\pm^{(p)}(\mathbf{q})\propto\left|\mathbf{q}\right|^{\frac{1}{n}}$ that the two EP$_{2n-1}$ can be decomposed into cycles  \cite{Moro1997,Ma1998,Demange2011}.
 One of these cycles is of order $n$ for each of them, yielding the $\left|\mathbf{q}\right|^{\frac{1}{n}}$ scaling of the dispersive bands, with the others remaining zero-energy FBs under the action of the perturbation term $H(\mathbf{q})$.

\subsection{Hofstadter spectrum and Landau levels of $n$-root Lieb lattice}

In this section, we consider the $n$-root Lieb lattice with uniform magnetic flux per plaquette.
As with the case of $n$-root graphene addressed in Sec.~\ref{subsec:hofstnrootgraph}, we take the uniform flux per plaquette to be in effect only in the inside of the plaquettes (white regions in Fig.~\ref{fig:nrootlieb}), \text{i.e.}, there is no added flux within the loop modules, such that they preserve their phase configuration.
This ensures that, upon raising the Hamiltonian of the periodic $n$-root Lieb lattice to the $n$th power, the diagonal block describing the blue SL$_1$ sites will correspond to the Lieb lattice with uniform flux per plaquette.
\begin{figure*}[ht]
	
	\begin{centering}
		\includegraphics[width=0.97 \textwidth]{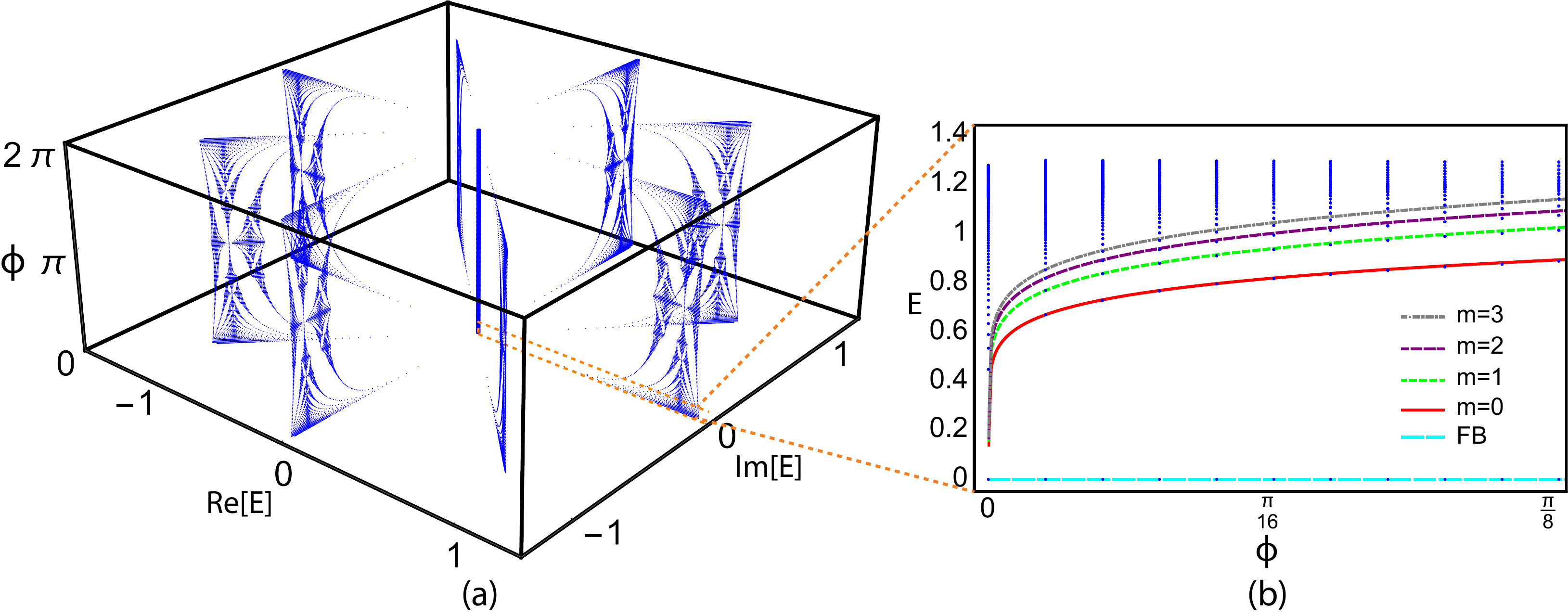}
		\par\end{centering}
	\caption{(a) Complex energy spectrum of the periodic 4-root Lieb lattice as a function of $\phi=2\pi\frac{l}{q}$, with $l=0,1,\dots,q$ and $q=157$. (b) Zoomed plot of of the low-flux region of the $p=0$ real energy branch in (a) delimited by the dashed orange box. The curves of different colors and formats represent the energy $E_{+m}^{(0)}(\phi)$, given in (\ref{eq:enerllsnrootliebdispersive}), of the corresponding LLs, except for the dashed cyan curve at zero energy that represents the degenerate LLs of the branch FBs, as per (\ref{eq:enerllsnrootliebflat}).}
	\label{fig:hofstspeclieb}
\end{figure*}

In Fig.~\ref{fig:hofstspeclieb}(a), we plot the complex Hofstadter spectrum for $n=4$ under PBC, with $N_x=3$ ($N_y=q=157$) unit cells in the $x$-direction ($y$-direction) where the four-branch structure can be clearly seen, since the flux does not break the generalized chiral symmetry $C_4$.
Raising this spectrum to the fourth power generates the fourfold degenerate Hofstadter spectrum of the Lieb lattice \cite{Aoki1996} (with three extra flux insensitive FBs at zero-energy due to SL imbalance).
Note that the Hofstadter spectrum for the Hermitian $2^n$-root versions of the square lattice, which has the Lieb lattice as its square-root model, was already addressed by some of the authors in a previous work \cite{Marques2023}.

Taking now $H_1(\mathbf{q})$ to be the low-energy expansion around the $\Gamma$ point of the Lieb lattice constructed with the blue SL$_1$ sites, ordered within the unit cell as described at the beginning of Sec.~\ref{sec:nrootlieb}, and applying the standard treatment to determine the LLs in the Landau gauge used in Sec.~\ref{subsec:hofstnrootgraph}, the energies and corresponding wavefunctions for the LLs of the Lieb lattice \cite{Goldman2011} can be determined to read as
\begin{widetext}
\begin{eqnarray}
	E_{\pm m}(\phi)&=&\pm v_{_F}\sqrt{(2m+1)\phi},
	\label{eq:enerlandaulieb}
	\\
	\ket{\psi_{\pm m}}&=&\frac{1}{\sqrt{4m+2}}
	\begin{pmatrix}
		\pm\sqrt{2m+1}\ket{m}
		\\
		-i\left(\sqrt{m}\ket{m-1}+\sqrt{m+1}\ket{m+1}\right)/\sqrt{2}
		\\
		\left(\sqrt{m}\ket{m-1}-\sqrt{m+1}\ket{m+1}\right)/\sqrt{2}
	\end{pmatrix},
	\label{eq:lleigsliebdispersive}
\end{eqnarray}
\end{widetext}
while the solutions for the FB are given by
\begin{widetext}
\begin{eqnarray}
	E_{0 m}(\phi)&=&0,
	\label{eq:enerlandauliebfb}
	\\
	\ket{\psi_{0 m}}&=&\frac{1}{\sqrt{4m+2}}
	\begin{pmatrix}
		0
		\\
		-i\sqrt{m+1}\ket{m-1}+i\sqrt{m}\ket{m+1}
		\\
		\sqrt{m+1}\ket{m-1}+\sqrt{m}\ket{m+1}
	\end{pmatrix}.
		\label{eq:lleigsliebflat}
\end{eqnarray}
\end{widetext}

The energy of the LLs of the $n$-root Lieb lattice can be directly determined from taking the $n$th root of (\ref{eq:enerlandaulieb}) and considering all branches in the complex energy spectrum, leading to
\begin{eqnarray}
	E_{\pm m}^{(p)}(\phi)&=&\pm\omega_n^{p/2}v_{_F}^\frac{1}{n}\left[(2m+1)\phi\right]^\frac{1}{2n},
	\label{eq:enerllsnrootliebdispersive}
	\\
	E_{0 m}^{(p)}(\phi)&=&0.
	\label{eq:enerllsnrootliebflat}
\end{eqnarray}
Fig.~\ref{fig:hofstspeclieb}(b) zooms in on the low-flux region of the real branch $p=0$ [highlighted in Fig.~\ref{fig:hofstspeclieb}(a)] of the 4-root Lieb lattice, where the curves following the analytical expression in (\ref{eq:enerllsnrootliebdispersive}) for the lowest LLs, and for all LLs in (\ref{eq:enerllsnrootliebflat}), are shown to be in good agreement with the numerical results.
The approximate solution for the eigenvectors of the dispersive bands at finite flux has the same form of  (\ref{eq:vecsnrootgraphlandau}), where the $h_j$ elements, with $j=1,\dots,n$, are now the quantized forms of the corresponding $h_j(\mathbf{q})$ elements of the $n$-root Lieb lattice.
As for LLs of the $n$-root graphene model in Sec.~\ref{subsec:hofstnrootgraph}, the approximate nature of these solutions comes from the fact only terms up to leading order in $|\mathbf{q}|$ are kept in $H_1(\mathbf{q})$, enabling the compact formula determined in (\ref{eq:lleigsliebdispersive}) and used in (\ref{eq:vecsnrootgraphlandau}).

Regarding the eigenvectors of the LLs of the zero-energy bands of the $n$-root Lieb lattice, (\ref{eq:vecsnrootgraphlandau}) cannot be used as it leads to divergences.
It can, however, point us in the direction of the solution, by noticing that the faster divergence occurs for the SL$_2$ component as $E\to0$, that is, one can expect the SL$_2$ component to be the dominant contribution of these states.
Indeed, it is possible to show that there is a \textit{single exact} solution at each LL that satisfies $H\ket{\phi_{0m}}=0$.
Since, for the $m$th LL,  this is the only solution for the FBs of all the $n$ branches (\textit{i.e.}, those not originating from sublattice imbalance), then $\ket{\phi_{_{\text{SL}_2,m}}}$ corresponds to an \textit{exceptional} LL (ELL) \cite{Bagarello2016} of at least order $n$ (since it may be increased by incorporating states from the sublattice imbalance originated subspace of $n-1$ FBs).
With $H$ the quantized form of (\ref{eq:hamiltnroot}) and assuming that $\ket{\phi_{0m}}=\left(\vec{0}_{_{\text{SL}_1}},\ket{\phi_{_{\text{SL}_2,m}}}^T,\vec{0}_{_{\text{SL}_3}},\dots,\vec{0}_{_{\text{SL}_n}}\right)^T$, where $\vec{0}_\mu$ is a zero vector with dimension $\mu$ and $\ket{\phi_{_{\text{SL}_2,m}}}$ is a four-component vector on the SL$_2$ subspace (there are four green SL$_2$ sites per unit cell in Fig.~\ref{fig:nrootlieb}), the eigenequation reduces to $h_1\ket{\phi_{_{\text{SL}_2,m}}}=0$, with
\begin{equation}
	h_1=\sqrt[n]{J}
	\begin{pmatrix}
		1-i\sqrt{\frac{B}{2}}\left(b+b^\dagger\right)&1&1&1+\sqrt{\frac{B}{2}}\left(b^\dagger-b\right)
		\\
		1&1&0&0
		\\
		0&0&1&1
	\end{pmatrix},
\end{equation}
whose solution is given by
\begin{widetext}
\begin{equation}
\ket{\phi_{_{\text{SL}_2,m}}}=\frac{1}{2\sqrt{2m+1}}
	\begin{pmatrix}
		-i\sqrt{m+1}\ket{m-1}+i\sqrt{m}\ket{m+1}
		\\
		i\sqrt{m+1}\ket{m-1}-i\sqrt{m}\ket{m+1}
		\\
		\sqrt{m+1}\ket{m-1}+\sqrt{m}\ket{m+1}
		\\
		-\sqrt{m+1}\ket{m-1}-\sqrt{m}\ket{m+1}
	\end{pmatrix}.
	\label{eq:lleigsnrootflat}
\end{equation}
\end{widetext}
Note that the precise order of these ELLs is difficult to extract from the Jordan decomposition of the Hofstadter spectrum, such as that of Fig.~\ref{fig:hofstspeclieb}(a), as errors from the numerical diagonalization of the Hamiltonian at a given $\phi$ infinitesimally lift the degeneracy of the $E=0$ states, thus breaking away from the condition for EP formation. 
At the same time, these ELLs have to be distinguished from the LLs studied in \cite{Zhang2020}, also labeled as exceptional.
There, the focus was on the Landau quantization around EPs (at zero flux), from which a real energy spectrum of \textit{regular} LLs was derived.
In our case, on the other hand, not only is the Landau quantization performed around an EP, but the resulting zero-energy level is \textit{itself} an EP, that is, an ELL.

\section{Photonic Ring Implementation of 3-root graphene} \label{sec:photonic}

We show here how photonic ring resonator systems provide a valid platform for the experimental probing of the discussed EPs. The implementation is based on a two-dimensional (2D) array of resonant rings, coupled through adequately engineered antiresonant link rings. First, a difference in optical path between upper and lower halves generates the phase component in the couplings that generates the energy downshift \cite{Hafezi2013}. Then, the required unidirectionality is produced via a distribution of gain and loss, modeled through a nonzero imaginary component of the refractive index, such that the upper half of the ring has a gain parameter $h$ and the lower half an equal amount of loss \cite{Longhi2015}. Effectively, this produces an asymmetry in the coupling strength $J_{-}/J_{+} = e^{2h}$ which approaches unidirectionality as $h$ increases, with $+\  (-)$ representing the forward (backward) direction. Perfect unidirectionality is only achieved in the limit of $h\to\infty$, and as such some deviations from the theoretical model are expected to appear. 

We consider planar rings with core refractive index $\tilde{n}_{\text{core}} = \num{3}$ surrounded by air and a width of $w=\SI{250}{\nano\meter}$, arranged as displayed in Fig.~\ref{fig:3rootgrapheucell}(b). The main rings have a radius of $R_M = \SI{6}{\mu\meter}$, while the link rings are constructed to fulfill the antiresonant condition $\beta(L_M-L_L) = (2\nu+1)\pi$ for a propagation constant $\beta$, and a main (link) ring circumference length of $L_M$ ($L_L$). The antiresonant condition may be fulfilled by several different ring sizes, depending on the value of $\nu$, and therefore we have the freedom to choose a smaller radius of $R_L = \SI{3.64}{\micro\meter}$ that reduces any possible direct coupling between link rings.
The imaginary part of the refractive index in their cores is then defined as $\text{Im}[\tilde{n}_{\text{link}}] = -0.1 \sin{\varphi}$, where $\varphi$ is the angle of the local polar coordinates of the ring, to achieve the split gain and loss distribution, which produces an asymmetry in the couplings of $\alpha \equiv J_{-}/J_{+} \approx 0.008$ \cite{Viedma2024}. We note that a sinusoidal pattern is in no way necessary as long as this asymmetry is produced, but a smooth profile is adopted to avoid reflection effects.
A constant relative distance between main and link rings of $d = \SI{0.56}{\micro\meter}$ is chosen, taken from the outer ring layers. 
Finally, the colored links in Fig.~\ref{fig:3rootgrapheucell}(a), wherein the couplings pick up a phase according to $J \to e^{\pm i 2\pi/3}\,J$, are produced by displacing the corresponding link rings from the center of the coupling line of the main rings by a distance of $d_y = \pm \SI{0.113}{\micro\meter}$ depending on the sign. With this, the required energy downshift is introduced, and one gains access to the EPs.
\begin{figure}[h]
	\begin{centering}
		\includegraphics[width=1\columnwidth]{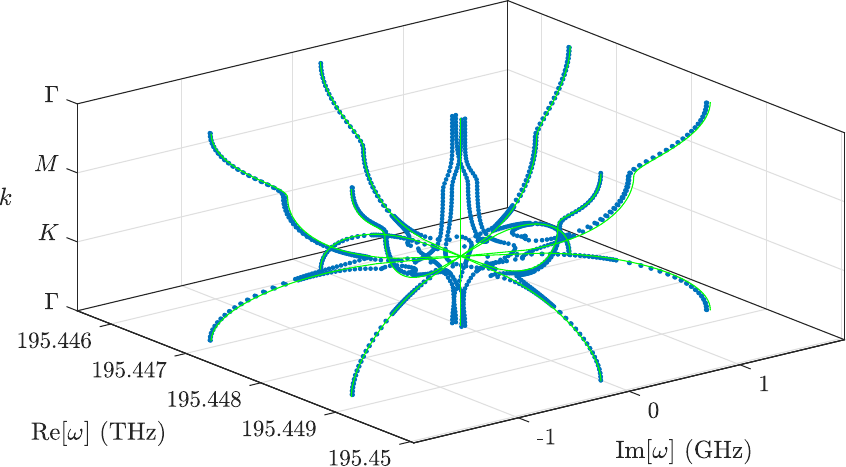} 
		\par\end{centering}
	\caption{Complex spectrum of eigenfrequencies for the cubic root of graphene with real phase factors under PBC, with the blue dots representing the finite-element simulation results and the light green lines the tight-binding predictions. A higher density of points was used near the EP at $K$.}
	\label{fig:photgraphene}
\end{figure}

We simulate such a system under PBC using the commercial finite-element simulation software COMSOL Multiphysics. The obtained eigenfrequency spectrum, which plays the role of the energy here, is displayed in Fig.~\ref{fig:photgraphene} as one goes along the high-symmetry line of the Brillouin zone. There, the simulated values are indicated by blue dots, while the overlapped light green lines are obtained by diagonalizing the tight-binding Hamiltonian in (\ref{eq:hamiltnroot}). One readily observes that the photonic ring implementation accurately reproduces the results away from the EP, including both the sublinear scaling of the energy indicated in (\ref{eq:enernrootgraphep}) and the splitting of the bands in branches that is characteristic of $n$-root TIs \cite{Marques2022,Viedma2024}. Nevertheless, a comparison with the tight-binding results reveals clear deviations in the inner bands, which are no longer flat. A similar effect appears when simulating the spectrum without the downshifting phase factors, as we showcase in Appendix~\ref{sec:app-3graphene}. There, for the same parameter values, deviation of the flat bands around zero energy also occurs, although it seems to be constant along the high-symmetry line as opposed to the nonuniform deformations showcased here. For completion, we also include in Appendix~\ref{sec:app-3lieb} the implementation of the 3-root Lieb model in photonic rings. Aside from a few differences, such as the need to use elongated rings for certain couplings due to geometrical constraints, the procedure to build the simulations is equivalent to the one explained here. The results are also in line with those for graphene: near-perfect agreement for the outer bands, and stronger deviations close to the EP.

We finish the discussion by noting that the eigensolver yields the eigenmodes for both ring circulations and, as such, points in the eigenfrequency plots are doubled. The fact that these points are not perfectly overlapped hints at small couplings between opposite ring circulations appearing within the same ring, which are usually neglected \cite{Hafezi2011}. Additionally, deviations from perfect unidirectionality in the ring couplings also play a role. These effects add to the known sensitivity of EPs \cite{Hodaei2017,Wiersig2020,Mao2024} and flat bands \cite{Viedma2024b,Riva2025} to perturbations, producing the band deformations shown in the photonic simulations. In Appendix~\ref{sec:app-deviations}, we provide an extended description of these extra couplings, and explain through tight-binding simulations how they qualitatively produce the observed deviations from the clean root model.

\section{Conclusions}
\label{sec:conclusions}

We outlined a method to generate $n$-root versions (with $n\geq 3$ an integer) of the parent hexagonal and Lieb lattices, based on connecting the basic constituents, namely, loop modules of unidirectional couplings \cite{Viedma2024}, in appropriate geometrical patterns.
The $n$-root models display a complex energy spectrum made of $n$ branches (with one of them purely real), plus extra zero-energy FBs induced by sublattice imbalance \cite{Marques2022}.
When no phases are included at the unidirectional couplings, the finite energy DP of the parent model translates as a DP at each of the branches of the $n$-root model, with a renormalized Fermi velocity for the associated Dirac cone.
Different behavior occurs, on the other hand, when the DP of the parent model is pushed to zero-energy by suitably tuning the phases configuration of the couplings.
In this scenario, the DP translates into zero-energy EPs of order $n$ (and higher) that are common to all branches, with the parent Dirac cone converted into an exceptional horn, characterized by sublinear dispersion relations in the neighborhood of the EPs and infinite Fermi velocity.

Applying a uniform transverse magnetic flux within the plaquettes formed by the loop modules of both $n$-root models generates a complex Hofstadter spectrum that, when raised to the $n$th power, recovers an $n$-fold degenerate Hofstadter spectrum of the parent model, and extra (flux-insensitive) FBs.
These originate from sublattice imbalance and are protected by the generalized chiral symmetry, which is not broken by the magnetic field.
When the phases configuration is engineered to remove the energy shift of the parent model, the LLs for the dispersive bands around the EP region \cite{Zhang2020} of each branch exhibit unusual energy scaling with both the magnetic flux $\phi$ and the LL number $m$.
More precisely, from the close relation between parent and respective $n$-root model, the energy of the LLs of the latter scales with the $n$th root of the corresponding LLs of the former.
For both graphene and the Lieb lattice as the parent model, this scaling thus becomes $\left| E_{\pm m}^{(p)}(\phi)\right|\sim \phi^{\frac{1}{2n}}\sim m^{\frac{1}{2n}}$ at every $p$ branch [see Eqs.~(\ref{eq:enerllsnrootgraph}) and (\ref{eq:enerllsnrootliebdispersive})].

To validate these results, we have built finite-element simulations for the root lattices of both graphene and the Lieb model in photonic ring resonators. The implementation is based on the usage of antiresonant link rings with a nonuniform distribution of optical gain and loss. These links enable near unidirectional effective couplings between the main rings of the lattice, and can allow fine-tuning of the coupling phases just by adjusting their positioning. The simulated spectra display good agreement with the theoretical results. One observes, however, significant deviations around the EPs, which extend to the degenerate FBs. 
The main cause of the deviations from the clean model is the existence of a coupling between opposite ring circulations, either due to reflection effects or a small direct coupling. Even if its value is small, flat bands and the EPs appear to be very sensitive to its effects, so focus should be placed on reducing it for future work. A possibility to mitigate these effects could be to introduce a bias that strongly separates the spectra for both ring circulations in energy, be that along the real (by detuning) or imaginary axes (through losses). An idea similar to Refs.~\cite{Ren2018,Hayenga2019,Liu2021NP} could be employed, wherein additional coupling elements are placed within the inner radii of the resonators to cause a strong suppression of one circulation via unidirectional couplings to the opposite one.

Several new directions remain open for exploration in the topic of $n$-root models.
In particular, unusual energy scaling is expected to be found, e.g., by taking the parent model to be (i) a Weyl semimetal with a linearly dispersive zeroth LL along the extra momentum coordinate \cite{Nielsen1983}, (ii) the Haldane model with linear chiral edge bands crossing the Fermi level in a ribbon geometry \cite{Marques2021b}, (iii) a semi-Dirac model where the LLs scale as $E\sim\phi^{\frac{2}{3}}$ \cite{Dietl2008}, or (iv) lattices hosting singular FBs, whose LLs scale with the LL number as $E\sim m^{-1}$ \cite{Rhim2021}.
It would also be interesting to study $n$-root lattices whose parent model is itself non-Hermitian and hosts EPs, such as graphene with an additional imaginary mass term \cite{Szameit2011,Ramezani2012}, where ``tachyonic'' modes with superluminal velocities appear in the vicinity of the EPs.
By constructing the $n$-root versions of such models, multiple concatenated mechanisms for EP generation will be present, and their alignment can lead to the amplification of the EP order, further enhancing its sensitivity to perturbations.
We are currently putting several of these predictions to the test, and their detailed analysis will be the subject of forthcoming studies.

\textit{Note added.} Recently, we became aware of two recent works: one \cite{McCann2026} with some related results based on a different approach that derives $n$-root versions of \textit{squared} parent models, and another \cite{Yoshida2026} that addresses the topology of Hopf EPs in higher-dimensional models.

\section{Acknowledgments}
 
A.M.M. and R.G.D. acknowledge funding by national funds through FCT - Funda\c{c}\~{a}o para a Ci\^{e}ncia e a Tecnologia, I.P., under the project UID/50025/2025 (doi.org/10.54499/UID/50025/2025), UID/PRR/50025/2025 (doi.org/10.54499/UID/PRR/ 50025/2025), UID/PRR2/50025/2025 (doi.org/10.54499/UID/PRR2/50025/2025) and the Associate Laboratory I3N - LA/P/0037/2020 (doi.org/10.54499/LA/P/0037/2020).
A.M.M. acknowledges financial support from i3N through the work Contract No.~CDL-CTTRI-91-SGRH/2024.
D.V. and V.A. acknowledge financial support from the Spanish Ministerio de Ciencia e Innovaci\'{o}n (MCIN) (MCIN/AEI/10.13039/501100011033, contracts No.~PID2020-118153GB-I00 and PID2024-160393NB-I00) and Generalitat de Catalunya (Contract No.~SGR2021-00138).
D.V. acknowledges funding from MCIN (MCIN/AEI/10.13039/501100011033, contract No.~PID2023-149988NB-C21) and funding from the European Union NextGenerationEU (PRTR-C17.I1).

\appendix

\section{Perturbation theory in $n$-partite lattices}
\label{app:perturbtheory}
In the following, we describe the application of perturbation theory to general $n$-partite lattices, with a focus on expanding around Dirac points (DPs) and exceptional points (EPs).
The results in this appendix come as a generalization to $n$-partite lattices of those in \cite{Yang2023}, where the technique for expanding around EPs in bipartite ($n=2$) lattices was developed.

We consider the bulk Hamiltonian of a general $d$-dimensional $n$-partite model of the form of (\ref{eq:hamiltnroot}), whose elements $h_j=h_j(\mathbf{k})$, where $\mathbf{k}$ is a  $d$-dimensional 
momentum vector, are rectangular matrices of size $d_j\times d_{j+1}$, with $d_j\in\mathbb{N}$ and $d_{n+1}\equiv d_1$.
From (\ref{eq:diagblocks}), we have that 
\begin{equation}
		H_1(\mathbf{k})=h_1(\mathbf{k})h_{2}(\mathbf{k})\dots h_{n}(\mathbf{k})=\prod\limits_{j=1}^n h_j(\mathbf{k}),
		\label{eq:h1app}
\end{equation}
which is a square matrix of size $d_1$ and $d_1\leq d_{j>1}$ is assumed without loss of generality, since all $H_j(\mathbf{k})$ diagonal blocks are isospectral modulo zero-energy flat bands (FBs).
From the generalized index theorem \cite{Marques2022}, the extra bands of larger blocks, in relation to $H_1(\mathbf{k})$, are precisely these FBs locked at zero energy, which we can therefore neglect in our perturbative analysis.
Diagonalization of $H_1(\mathbf{k})$ then yields $d_1$ energy bands labeled as $E_{1,l}(\mathbf{k})$, with $l=1,2,\dots,d_1$.
Similarly to (\ref{eq:ebandsnrootgraph}), the energy spectrum of the original $n$-partite model defined by $H(\mathbf{k})$, modulo zero-energy FBs, is given by
\begin{equation}
	E_{l}^{(p)}(\mathbf{k})=\omega_n^{p} E_{1,l}^{\frac{1}{n}}(\mathbf{k}),	
	\label{eq:ebandsnrootgen}
\end{equation}
with $\omega_n=e^{i\frac{2\pi}{n}}$ and $p=0,1,\dots,n-1$ is the branch index.

Suppose we want to expand $H(\mathbf{k})$ around some arbitrary point $\mathbf{k_*}$ from which the momentum is measured as $\mathbf{q}\equiv\mathbf{k}-\mathbf{k_*}$, which implies expanding its constituent nontrivial elements $h_j(\mathbf{k})$.
Even though we cut off our expansion at first order in $\mathbf{q}$, our approach can be readily generalized to arbitrary order.
The expanded elements read as
\begin{eqnarray}
h_j(\mathbf{q})&\approx& h_j(\mathbf{k_*})+v_j(\mathbf{q}),
\label{eq:hjapp}
\\
v_j(\mathbf{q})&=&\nabla_{\mathbf{k}}h_j(\mathbf{k})\vert_{\mathbf{k}=\mathbf{k_*}}\cdot\mathbf{q}.
\end{eqnarray}
The expansion of $H_1(\mathbf{k})$ in (\ref{eq:h1app}) becomes, with the use of (\ref{eq:hjapp}), 
\begin{equation}
H_1(\mathbf{q})=\prod\limits_{j=1}^n h_j(\mathbf{q})=\prod\limits_{j=1}^n \left(h_j(\mathbf{k_*})+v_j(\mathbf{q})\right),
\end{equation}
which generates terms up to $n$th order in $|\mathbf{q}|$.
Keeping only the terms to leading order yields
\begin{eqnarray}
H_1(\mathbf{q})&\approx& H_1(\mathbf{k_*})+V_1(\mathbf{q}),
\label{eq:h1appexpanded}
\\
H_1(\mathbf{k_*})&=&\prod\limits_{j=1}^n h_j(\mathbf{k_*}),
\label{eq:h1appatkc}
\\
V_1(\mathbf{q})&=&\sum\limits_{j=1}^n A_j(\mathbf{q}),
\end{eqnarray}
where $A_j(\mathbf{q})$ is the same matrix product as in (\ref{eq:h1appatkc}), but with $v_j(\mathbf{q})$ in place of $h_j(\mathbf{k_*})$,
\begin{equation}
	A_j(\mathbf{q})=h_1(\mathbf{k_*})h_2(\mathbf{k_*})\dots v_j(\mathbf{q})\dots h_{n-1}(\mathbf{k_*})h_n(\mathbf{k_*}).
\end{equation}

At this point, we restrict our analysis to the relevant cases of this study, namely, to $d_1=2,3$ and a single \textit{untilted}, \textit{massless}, and \textit{symmetric} pseudospin-$\frac{d_1-1}{2}$ cone in the spectrum of a \textit{Hermitian} $H_1(\mathbf{q})$ block.
Under these assumptions, diagonalization of $H_1(\mathbf{q})$ in (\ref{eq:h1appexpanded}) yields energy bands with linear dispersion of the form
\begin{equation}
	E_{\pm}(\mathbf{q})=E_0\pm v_{_F}|\mathbf{q}|,\ \ \ \ \text{for}\  d_1=2,3,
	\label{eq:ebandsh1app}
\end{equation}
where $E_0$ is a constant energy shift and $v_{_F}$ is the Fermi velocity.
For $d=3$, an extra flat band at $E_0(\mathbf{q})=E_0$ completes the spectrum.
Finally, the energy bands of $H(\mathbf{q})$ are again obtained from (\ref{eq:ebandsh1app}) as
\begin{equation}
	E_{\pm}^{(p)}(\mathbf{q})=w_n^pE_{\pm}^{\frac{1}{n}}(\mathbf{q})=w_n^p \left(E_0\pm v_{_F}|\mathbf{q}|\right)^{\frac{1}{n}},
	\label{eq:ebandsapphq}
\end{equation}
together with $E_0^{(p)}=w_n^pE_0^{\frac{1}{n}}$ for $d_1=3$.
The behavior of the dispersive bands around $\mathbf{k}_*$ depends on the value of $E_0$:
\begin{itemize}
	\item[(i)] 
	If $E_0\neq 0$, then the expression in (\ref{eq:ebandsapphq}) has to be further expanded around $\mathbf{q}=\mathbf{0}$, yielding, to leading order,
	\begin{equation}
		E_{\pm}^{(p)}(\mathbf{q})\approx w_n^p\left[E_0^{\frac{1}{n}}\pm v_{_F}^\prime |\mathbf{q}|\right],
	\end{equation}
	with a renormalized Fermi velocity
	\begin{equation}
			v_{_F}^\prime=\frac{1}{n}E_0^{\frac{1-n}{n}}v_{_F}.
	\end{equation}
	The complex linear dispersion on $ |\mathbf{q}|$ means that there is an $n$-tuple of DPs for the $n$-root model at $\mathbf{k_*}$, which translates as a DP in the spectrum of each of the $n$ diagonal blocks $H_j(\mathbf{k})$.  
	
	\item[(ii)] 
	If $E_0=0$, then (\ref{eq:ebandsapphq}) reduces to
	\begin{equation}
		E_\pm^{(p)}(\mathbf{q})=\pm w_n^{p/2} \left(v_{_F}|\mathbf{q}|\right)^{\frac{1}{n}},	
	\end{equation}
	where the bands were relabeled to ensure that each branch is defined by a unique pair of symmetric bands in the complex energy spectrum.
	The complex sublinear dispersion on $|\mathbf{q}|$ means that there are two EPs for the $n$-root model at $\mathbf{k_*}$ and zero-energy, which translates as a zero-energy DP in the spectrum of each of the $n$ diagonal blocks $H_j(\mathbf{k})$. The EPs are of order $n$ or higher. The latter case relates to the extra zero-energy FBs that appear due to sublattice imbalance \cite{Marques2022}, whose corresponding eigenvectors also coalesce to one of the EP eigenvectors at $k_*$, therefore increasing its order, as shown for the models studied in the main text. 
\end{itemize}

\section{Photonic 3-root graphene without energy downshift} \label{sec:app-3graphene}

\begin{figure}[h]
	\begin{centering}
		\includegraphics[width=1\columnwidth]{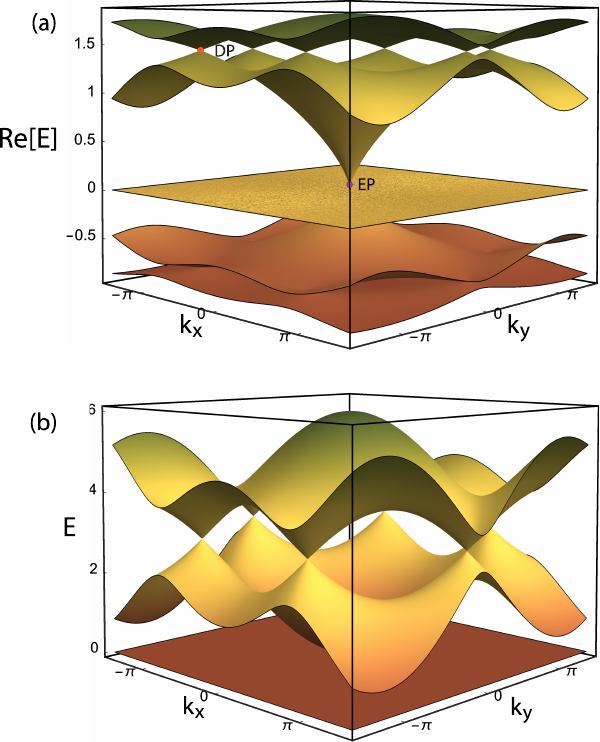} 
		\par\end{centering}
	\caption{(a) Real part of the bulk energy spectrum for the 3-root graphene model without downshifting phase terms. We highlight a DP at the $K$-point (orange dot), and an EP at the $\Gamma$ point (purple dot). (b) Energy spectrum of the cubed model of the Hamiltonian in (a), where the dispersive bands display an energy shift with respect to the bands of regular graphene. Here, the corresponding  DPs are located at $E=3J$.}
	\label{fig:graphene_nophase_3D}
\end{figure}

In Fig.~\ref{fig:graphene_nophase_3D}(a), we showcase the 3-root graphene model when the phase terms in Fig.~\ref{fig:3rootgrapheucell}(a) are not included in the couplings, thus corresponding to case (i) of App.~\ref{app:perturbtheory}. In this case, all the DPs at the $K$-points sit at nonzero energy, and therefore do not achieve sublinear energy scaling. Looking at the cubed spectrum in Fig.~\ref{fig:graphene_nophase_3D}(b), from which these points originate, we see that all dispersive bands are moved up in energy, with the DPs occurring at $E=3J$ as justified in Eq.~(\ref{eq:lowenergygraph}) in the main text. Notably, in this parent model we see a quadratic band touching between the bottom band and the FB. As a consequence, an EP appears in the root spectrum around $E=0$ as highlighted in Fig.~\ref{fig:graphene_nophase_3D}(a). Due to this quadratic origin, the low-energy behavior of the EP goes as $E\propto|\bf{q}|^{\frac{2}{3}}$, in contrast to the $|\bf{q}|^{\frac{1}{3}}$ scaling of the EPs described in the main text.

\begin{figure}[h]
	\begin{centering}
		\includegraphics[width=1\columnwidth]{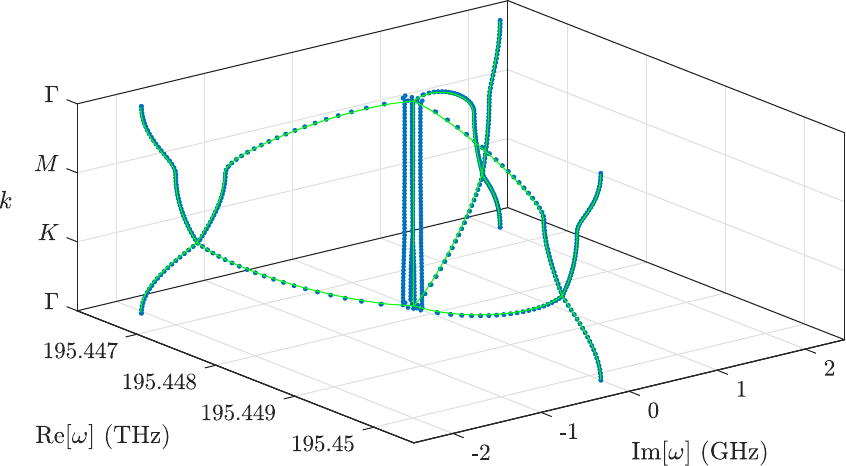} 
		\par\end{centering}
	\caption{Complex spectrum of eigenfrequencies for the 3-root of graphene under PBC, with the blue dots representing the finite-element simulation results and the light green lines the tight-binding predictions.}
	\label{fig:graphene_nophase}
\end{figure}

We now consider the photonic simulations of this root model, where the only difference to the ones in the main text is the absence of the perpendicular link ring displacements that generate the phase factors. In Fig.~\ref{fig:graphene_nophase}, we display the complex eigenfrequency spectrum along the high-symmetry line, where one readily observes that the DPs at the $K$ point are very accurately reproduced. Additionally, the zero-energy EP present at the $\Gamma$ point very clearly appears as well. 
More concretely, the eigenvector structure at $E(\Gamma)=0$ consists of two EP$_{n-1}$ plus a regular point, with the latter coupling to the former two in first-order in $\mathbf{q}$, leading to the low-energy $E\propto|\bf{q}|^{\frac{2}{3}}$ scaling for the dispersive bands.
From Fig.~\ref{fig:graphene_nophase}, one readily sees that the only significant distortion occurs for the flat bands around zero energy, which experience a small and nearly constant splitting across the $\mathbf{k}$-path.

\section{Photonic 3-root Lieb lattice} \label{sec:app-3lieb}

\begin{figure}[h]
	\begin{centering}
		\includegraphics[width=0.48 \textwidth]{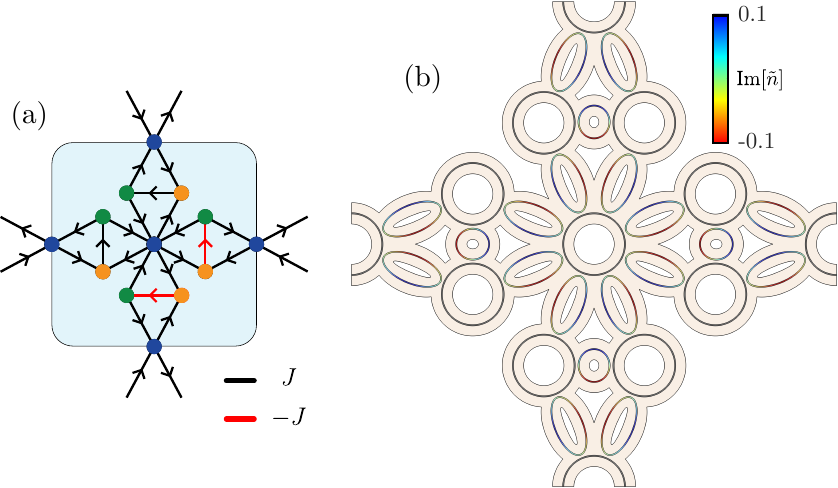}
		\par\end{centering}
	\caption{(a) Unit cell of the 3-root Lieb model with a modified hopping phase configuration, where red hoppings carry a $\pi$ phase factor, that is, an opposite coupling sign. (b) Photonic ring implementation of the unit cell shaded in blue in (a). Due to geometric constraints, link rings with two different shapes are considered. Equal coupling strengths are achieved for both types of rings by using different relative ring distances.}
	\label{fig:3rootLiebucell}
\end{figure}

\begin{figure}[t]
	\begin{centering}
		\includegraphics[width=1\columnwidth]{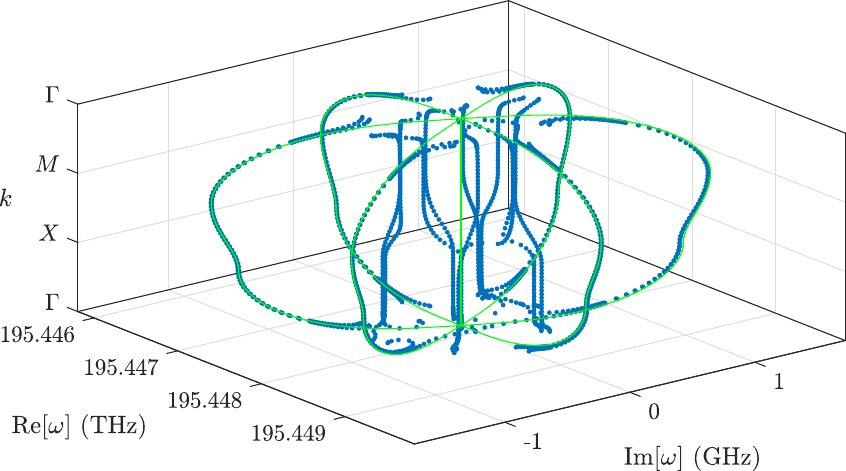} 
		\par\end{centering}
	\caption{Complex spectrum of eigenfrequencies for the 3-root of the Lieb model with the energy downshifting phase configuration under PBC, where the blue dots represent the finite-element simulation results and the light green lines the tight-binding predictions. The density of points is higher around the EP at the $\Gamma$ point.}
	\label{fig:photonLieb}
\end{figure}

The photonic ring implementation of the 3-root Lieb model follows the same ideas explained in Sec.~\ref{sec:photonic} of the main text and uses the same parameter values, with a key difference: looking at Fig.~\ref{fig:3rootLiebucell}(a), it is clear that some sites to eight coupling lines, in contrast to the six for graphene sites, forcing link rings to be more closely packed together. This imposes some geometrical constraints that are navigated by employing elongated rings of elliptical shape, as can be observed in Fig.~\ref{fig:3rootLiebucell}(b). Long links have semiaxis lengths $R_{a1} = \SI{6}{\micro\meter}$ and $R_{b1} = \SI{2.66}{\micro\meter}$, while short links have $R_{a2} = \SI{3.13}{\micro\meter}$ and $R_{b2} = \SI{2.94}{\micro\meter}$, both engineered to be antiresonant to the main rings. Following the coupling scheme in Fig.~\ref{fig:3rootLiebucell}(a), long link rings are placed at a distance of $d_{L} = \SI{0.54}{\micro\meter}$, and short links are instead placed at a distance of $d_{S} = \SI{0.529}{\micro\meter}$ to have approximately the same coupling strength. One should note, however, that deviations from this value do not significantly alter the results. The $\pi$ phase factor is achieved by displacing the corresponding short rings by a distance of $d_y = \SI{0.17}{\micro\meter}$ perpendicularly to the coupling line between the main rings.

For simplicity, we consider equal gain and loss distributions across all link rings, and of the same form as the 3-root graphene implementation. Since the nonreciprocity parameter $\alpha \equiv J_{-}/J_{+}$ depends both on the size of the rings as well as on the coupling strength \cite{Viedma2024}, this leads to slightly uneven $\alpha$ parameters. All of them, however, have been computed to be approximately equal to $\alpha \approx 0.007$.

Considering all of this, we simulate the photonic 3-root Lieb lattice under PBC and display the obtained spectra along the high-symmetry line in Fig.~\ref{fig:photonLieb}. Following the same behavior as the results for graphene, one observes virtually no deviations for the outer bands away from the EP at the $\Gamma$-point. On the other hand, the central FBs have a large constant deviation, which separates them radially from the center, as well as a smaller $k$-dependent variation.

\section{Characterizing the deviations in the photonic bands} \label{sec:app-deviations}

\begin{figure*}[t]
	\begin{centering}
		\includegraphics[width=1\textwidth]{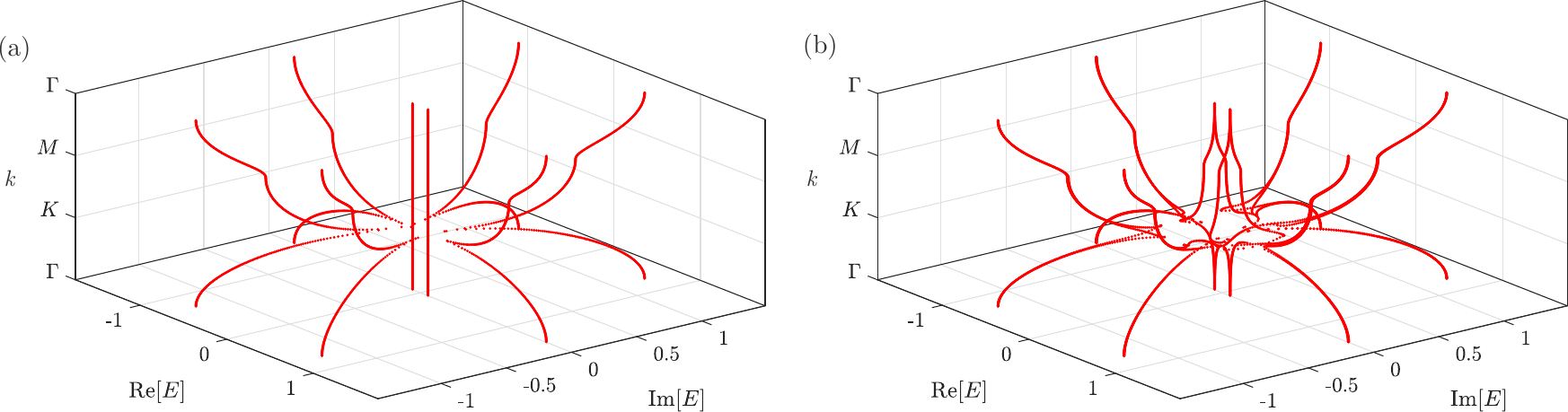} 
		\par\end{centering}
	\caption{(a) Energy spectrum for the tight-binding Hamiltonian of 3-root graphene, including the relevant phase factors, when a small backwards coupling $J_{-} = \alpha J$ with $\alpha = \num{8e-3}$ is considered. (b) Spectrum for the total Hamiltonian, considering both ring circulations, for a cross-circulation coupling of $J_{\text{cc}}=\num{4e-3}$ and the same $\alpha$ as (a).}
	\label{fig:graphene_deviations}
\end{figure*}

We attribute the deviations from the tight-binding model to two main sources. First, the unidirectionality of the couplings is only approximate. In this system there always is a small backward component that can be modeled with the nonreciprocity parameter $\alpha$ by modifying the Hamiltonian so that $H\to H+\alpha H^\dagger$ since $H^\dagger$ yields the same model only with reversed coupling directions \cite{Viedma2024}. Computing the bands of the modified system reveals that small increases to $\alpha$ mainly contribute to splitting the flat bands along the real energy axis, see Fig.~\ref{fig:graphene_deviations}(a) for $\alpha = \num{8e-3}$. Clearly, in the limit of $\alpha\to 1$ we recover the purely Hermitian case and all bands collapse to the real energy axis.

\begin{figure*}[t]
	\begin{centering}
		\includegraphics[width=1\textwidth]{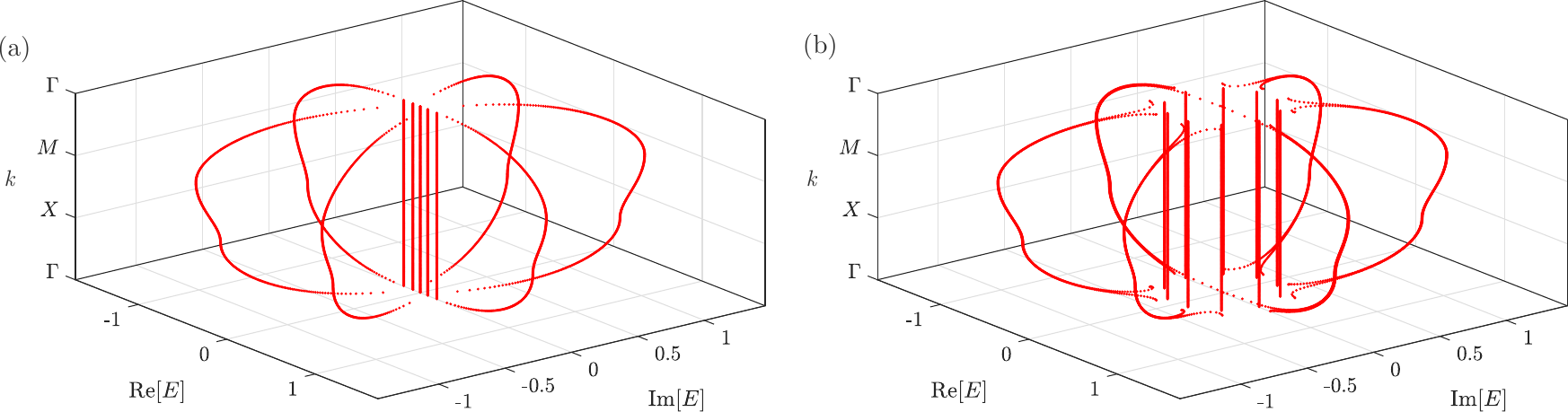} 
		\par\end{centering}
	\caption{(a) Energy spectrum for the tight-binding Hamiltonian of the 3-root Lieb model, including the relevant phase factors, with a small backward coupling determined by $\alpha \approx 0.007$. (b) Spectrum for the total Hamiltonian, considering both ring circulations, for a cross-circulation coupling of $J_{\text{cc}}=\num{4e-3}$ and the same $\alpha$ parameters as (a).}
	\label{fig:lieb_deviations}
\end{figure*}

The second source of discrepancy can be attributed to small coupling between opposite circulations in the rings. To treat this, we consider an enlarged Hilbert space containing both ring circulations. Reversing the circulation direction in the ring system implies a reversal of the coupling direction as well as of the sign of the real phases. For a phase factor of $\phi$, this is equivalent to the following change: $H(\phi)\to H^\dagger(-\phi)$. Therefore, the \textit{total} Hamiltonian in the enlarged Hilbert space reads as
\begin{equation}
    H_T = \begin{pmatrix}
        H(\phi) + \alpha H^\dagger(\phi) & 0 \\
        0 & H^\dagger(-\phi) + \alpha H(-\phi)
    \end{pmatrix}, \label{eq:nonzeroalpha}
\end{equation}
considering that both circulations produce their own independent Hamiltonians. Nevertheless, cross-circulation couplings must be included for a correct description of the system around the zero-energy region. These couplings can be caused either by reflections within the system, or by direct coupling to the opposite circulation across different main rings. This latter term can be justified by considering that main and link rings are not exactly parallel to one another in the coupling region due to their curvature. Although terms of higher complexity might be considered, we assume that the main deviations in the simulations can be modeled in the tight-binding model by a Hermitian coupling term $J_{\text{cc}}$ between opposite circulations within each main ring. These appear in the off-diagonal square blocks of Eq.~(\ref{eq:nonzeroalpha}) as $H_T(j,j+N)=H_T(j+N,j) = J_{\text{cc}}$ with $j=1, 2, \ldots, N$ and $N=8$ the number of sites in the unit cell.

In Fig.~\ref{fig:graphene_deviations}(b), we plot the energy bands for the tight-binding model for $\alpha = \num{8e-3}$ and $J_{\text{cc}} = \num{4e-3}$. Even if the values are small, the combination of these two effects produces non-trivial variations of the inner energy bands, especially around the EPs, while the outer bands are only slightly modified. Comparing with the finite-element simulations in Fig.~\ref{fig:photgraphene} in the main text, one observes that the deformations of the bands are qualitatively reproduced by the modified tight-binding model. Clearly, achieving a precise simulation of the model's spectrum is challenging physically due to the sensitivity of the flat bands and especially the EPs to the parameters discussed here. Still, the main characteristics of the system and the root scaling of the energy are correctly reproduced by the ring lattice.

We repeat the same calculations for the 3-root Lieb model in Figs.~\ref{fig:lieb_deviations}(a) and (b) for the single-circulation and total Hamiltonians, respectively. In this case, we disregard the minor differences in $\alpha$ for the couplings corresponding to different kinds of rings, and take a constant $\alpha \approx 0.007$. Interestingly, in this case even if the band deviations are somewhat strong, these effects produce mostly constant splittings of the flat bands, in contrast to the results in Fig.~\ref{fig:graphene_deviations}(b). We attribute the additional deformations observed in Fig.~\ref{fig:photonLieb} beyond this constant splitting to different loss terms due to the non-circular shape of the different link rings. These extra losses translate into variations of the effective model, both at the level of couplings and of imaginary onsite terms.

Regarding the tolerance of the main characteristics of the model to these perturbations, we start to observe significant deformations of the exceptional horns and the flat bands above $\alpha=0.1$ and $J_{cc} =0.01$. Therefore, we consider these orders of magnitude as approximate ceilings of acceptable perturbations, which are consistent with the values obtained from our photonic simulations. Nevertheless, we also note that other properties of the models survive for larger perturbations, such as the branching of bands governed by an approximate generalized chiral symmetry.

\bibliography{ehorns}

\end{document}